\documentclass[useAMS,usenatbib]{mn2e}
\usepackage{graphicx}
\usepackage[totalwidth=480pt, totalheight=680pt]{geometry}

\newcommand{\kmpers}{\mathrm{km s}^{-1}} 
\newcommand{\perpix}{\mathrm{pixel}^{-1}}
\newcommand{\muem}{\mu\mathrm{m}}
\newcommand{\bJ}{\mathrm{b}_{\mathrm{J}}}
\newcommand{\rF}{\mathrm{r}_{\mathrm{F}}}
\newcommand{\afe}{[\alpha/\mathrm{Fe}]}
\newcommand{\hbeta}{\mathrm{H}\beta} 
\newcommand{\hbetag}{\mathrm{H}\beta_\mathrm{G}} 
\newcommand{\mgb}{Mg\textit{b}} 

\title[The stellar populations of early-type galaxies -- I]{The
  stellar populations of early-type galaxies -- I. Observations,
  line-strengths and stellar population parameters}
\author[Harrison et al.]{Craig D. Harrison$^{1,2}$\thanks{E-mail:
charrison@ctio.noao.edu}, Matthew Colless$^{3}$, Harald
  Kuntschner$^{4}$, Warrick
  J. Couch$^{5}$, \newauthor Roberto De Propris$^{2}$, Michael B. Pracy$^{5}$\\
$^{1}$Research School of Astronomy \& Astrophysics, Australian
  National University, Weston Creek, ACT 2611, Australia\\
$^{2}$Cerro Tololo Inter-American Observatory, Casilla 603, La Serena,
  Chile\\
$^{3}$Anglo-Australian Observatory, PO Box 296, Epping, NSW 2121, Australia\\
$^{4}$Space Telescope European Coordinating Facility, European
  Southern Observatory, Karl-Schwarzschild-Str, 85748 Garching, Germany\\
$^{5}$Centre for Astrophysics \& Supercomputing, Swinburne University
  of Technology, PO Box 218, Hawthorn, VIC 3122, Australia}
\begin{document}

\date{Accepted ... . Received ... ; in original form 2009 December 15}

\pagerange{\pageref{firstpage}--\pageref{lastpage}} \pubyear{2010}

\maketitle

\label{firstpage}

\begin{abstract}
The influence of environment on the formation and evolution of
early-type galaxies is, as yet, an unresolved issue. Constraints can
be placed on models of early-type galaxy formation and evolution by
examining their stellar populations as a function of environment. We
present a catalogue of galaxies well suited to such an
investigation. The magnitude-limited ($\bJ\lid 19.45$) sample was
drawn from four clusters (Coma, A1139, A3558, and A930 at $\left< z
\right>=0.04$) and their surrounds. The catalogue contains
luminosities, redshifts, velocity dispersions and Lick line strengths
for 416 galaxies, of which 245 are classified as
early-types. Luminosity-weighted ages, metallicities, and
$\alpha$-element abundance ratios have been estimated for 219 of these
early-types. We also outline the steps necessary for measuring
fully-calibrated Lick indices and estimating the associated stellar
population parameters using up-to-date methods and stellar population
models. In a subsequent paper we perform a detailed study of the
stellar populations of early-type galaxies in clusters and investigate
the effects of environment.
\end{abstract}


\section{Introduction}\label{intro}

The stellar populations of all but the nearest galaxies are currently
unresolvable. Therefore their study relies on the analysis of their
integrated light. By examining the absorption lines in an integrated
spectrum one can estimate relative luminosity-weighted mean stellar
population parameters such as age, metallicity ([Z/H]), and
$\alpha$-element abundance ratio ($\afe$). The Lick/IDS group defined
a system of absorption-line indices that can be used as stellar
population tracers \citep{burstein84, faber85, burstein86, gorgas93,
  worthey94a, trager98}. These indices can be measured in a consistent
way and easily compared to predictions from single stellar population
(SSP) models. Recently, synthetic spectra, for the which the stellar
population parameters are known, have been used for this purpose
\citep[e.g.][]{vazdekis10}. When combined with improved methods for
estimating $\hbeta$ emission \citep{sarzi06}, this appears to be a
promising way forward.

Absolute determinations of the stellar population parameters are
complicated by the similarity of the effects of age and [Z/H] on a
galaxy's spectral energy distribution, the age--[Z/H] degeneracy
\citep{worthey94b}, and severely hampered by the effects of the
non-solar abundance ratios in early-type galaxies \citep{worthey96}.

\begin{table*}
\begin{minipage}{118mm}
\caption{Details of the clusters from which the galaxy samples were
drawn.}
\begin{tabular}{lcccr@{}lr@{}lcc}
\hline
Name & R.A. & Dec. & $z$ & \multicolumn{2}{c}{$\sigma_{clus}$} &
\multicolumn{2}{c}{$N_z$} & $R_A$ & B--M\\
\hline
Coma   & 12 59 48.7 &     $+27$ 58 50 & $0.0232 \pm 0.0002$ & 100
       & $8 \pm 33$  & 49 & 9 & 2  &      II\\ 
A1139  & 10 58 11.0 &     $+01$ 36 16 & $0.0396 \pm 0.0002$ &  50 
       & $4 \pm 47$  & 10 & 6 & 0  &     III\\
A3558  & 13 27 54.0 &     $-31$ 29 30 & $0.0480 \pm 0.0003$ &  97 
       & $7 \pm 37$  & 34 & 1 & 4  &       I\\
A930  & 10 07 01.3 &     $-05$ 37 29 & $0.0578 \pm 0.0003$ &  90
       & $7 \pm 82$  &  9 & 1 & 1  &     III\\
\hline
\end{tabular}
\label{clus details}
\medskip

Coordinates are given in J2000.0. The units of $\sigma_{clus}$ are
$\kmpers$; $N_z$ is the number of galaxies used to calculate the
cluster $z$ and $\sigma_{clus}$.
\end{minipage}
\end{table*}

It is possible to break the age--[Z/H] degeneracy through the use of
appropriate absorption-line indices, i.e.\ by combining an index more
sensitive to age variations with one more sensitive to [Z/H]
variations \citep{rabin82, gonzalez93}. By comparing the
line-strengths of two (or more) of these indices to predictions from
an SSP model through the use of line-diagnostic diagrams, stellar
population parameters can be estimated for individual galaxies. It
must be stressed that these are relative, luminosity-weighted mean
stellar population parameters; a small percentage of recently-formed
luminous stars can contribute a disproportionate amount of flux to an
integrated spectrum.

In luminous early-type galaxies, it is well known that Mg is observed
to be over-abundant relative to Fe when compared with the solar ratio
\citep[and others]{peletier89, worthey92a, davies93, fisher95,
  greggio97, mehlert98, jorgensen99a, kuntschner00}. This leads to
indices such as \mgb\ and Mg$_2$ yielding higher metallicities and/or
younger ages than indices such as Fe5270 and Fe5335, when compared to
stellar population models based on solar abundances. Every stellar
population model that uses Milky Way-based index calibrations suffers
from this bias in $\afe$ where indices reflect super-solar $\afe$ at
sub-solar metallicities \citep{borges95}. The current generation of
stellar population models, such as those of \citet{thomas03a} that are
used in this study, attempt to correct for this bias.

These models are based on the SSP models of \citet[][see also
  \citealt{maraston98}]{maraston02} and provide line-strengths for the
entire set of Lick indices. These models contain variable abundance
ratios with $[\alpha/\mathrm{Fe}]=0.0$, 0.3, 0.5 dex,
$[\alpha/\mathrm{Ca}]=-0.1$, 0.0, 0.2, and 0.5 dex, and
$[\alpha/\mathrm{N}]=-0.5$ and 0.0 dex. [$\alpha$/Fe] is particularly
important since it provides information on the formation time-scale of
the stellar population. The models cover ages between 1 and 15\,Gyr,
metallicities between 0.005 and 3.5 times solar, and are calibrated
with Milky Way globular clusters for which the metallicity and
[$\alpha$/Fe] are known from independent spectroscopy of individual
stars \citep{puzia02}.

Since SSP models with variable abundance ratios became available,
iterative methods have been devised to take into account the $\afe$ of
the stellar population when deriving its age and [Z/H] from
line-diagnostic diagrams. A more straight-forward method involves the
simultaneous fitting of a number of indices to the predictions from a
stellar population model using a $\chi^2$ technique
\citep*{proctor04}. The advantage of this method over the
line-diagnostic diagram is that the errors on each index can be taken
into account. Using a large number of indices maximises the use of the
available data and minimises the impact on derived parameters of
imperfect calibration to the Lick system and reduction errors. It is
this method that we use to estimate the stellar population parameters
for our galaxies.

The outline of the remainder of the paper is as follows. Sample
selection, observations and basic reductions are detailed in \S
\ref{obs and red}. Redshift ($z$) and velocity dispersion ($\sigma$)
measurements, and their errors, are discussed in \S \ref{red and
  sig}. \S \ref{indices} goes through the steps necessary to measure
the strength of absorption lines, fully calibrated to the Lick system,
including broadening the spectra and correcting the indices for
velocity dispersion broadening. The spectral classification of the
galaxies is detailed in \S \ref{detect sf}. Estimation of the stellar
population parameters and the effects on these estimates of non-solar
abundance ratios are discussed in \S \ref{parameters}. A summary of
the final catalogue and its potential uses are provided in \S
\ref{summary}.

A Hubble parameter $H_0=70 \mathrm{km s}^{-1} \mathrm{Mpc}^{-1}$,
matter density parameter $\Omega_M=0.3$ and dark energy density
parameter $\Omega_\Lambda=0.7$ are adopted throughout this work.

\section{Observations and Data Reductions}\label{obs and red}

\subsection{Sample selection}\label{sample}

The sample of galaxies observed were drawn from four clusters and
their surrounds. The clusters, Coma, A1139, A3558, and A930 with $0.02
\la z \la 0.06$, were selected to span the range of Abell richness
($R_A$) classes and Bautz--Morgan (B--M) classifications. Coma was
also selected to calibrate our data to the Lick system, since it has
been studied extensively and has published Lick index line-strengths
for its early-type galaxies. Details of the clusters are given in
Table \ref{clus details}.

A list of potential targets was created in the following
manner. Positions, spectroscopic redshifts, and magnitudes for all
galaxies inside the instruments' fields of view were obtained: for
A930 and A1139 these data were obtained from the 2dFGRS database
\citep{colless01, colless03}, while for A3558 they were obtained from
NED for the 2dF observations and from the 6dFGS source catalogue
\citep{jones04} for the 6dF observations. $\bJ$ and $\rF$ magnitudes
were obtained for the galaxies selected from NED by cross-correlating
with the 6dFGS source catalogue. The magnitude limit of 2dFGRS is $\bJ
= 19.45$ mag and so we impose this limit on the entire sample. The
data on the Coma galaxies were obtained from a catalogue compiled by
\citet{moore02}. There are 135 early-type galaxies in the
\citeauthor{moore02}\ catalogue, all located within the inner $\sim 1$
Mpc of the cluster. The B-band magnitudes were converted to $\bJ$
magnitudes using the equation found in \cite{blair82}.

The 6dF fields for A1139 and A930 were not centred on these clusters
because they are too close to the edge of the 2dFGRS survey
area. Instead, the cluster centre was positioned on one side of the
6dF field, which allowed galaxies at larger cluster-centric radial
distances to be observed.

\begin{figure*}
  \includegraphics[width=0.9\textwidth]{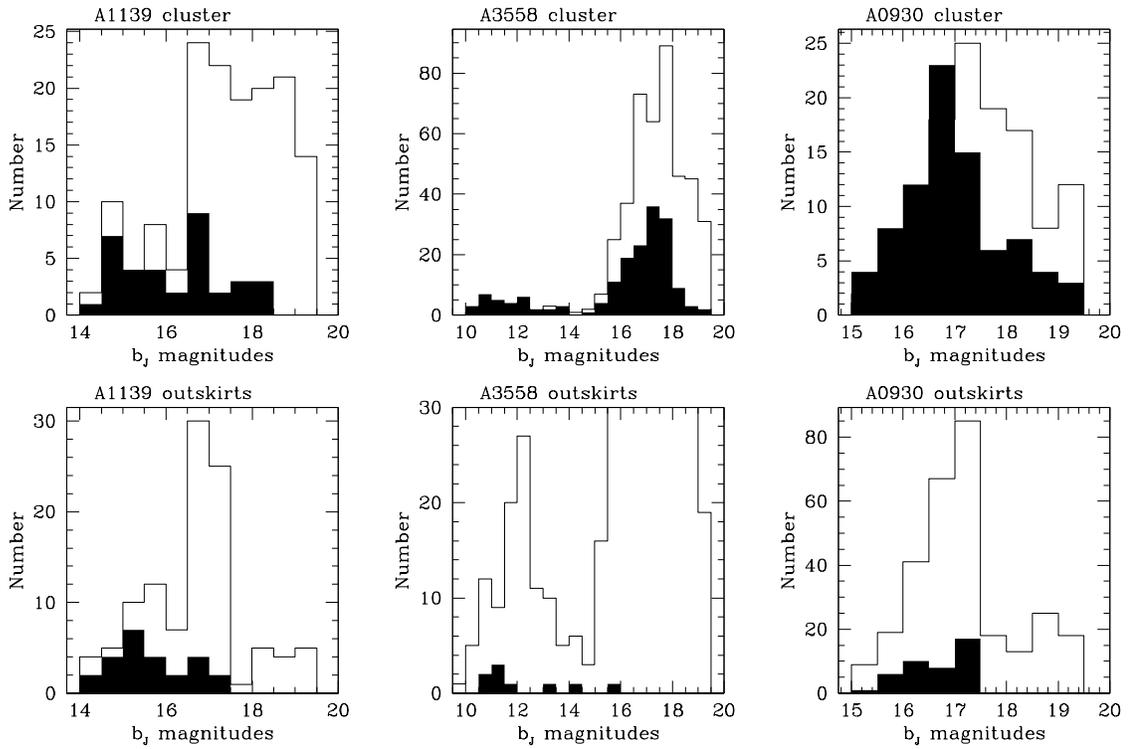}
  \caption{The number of galaxies, as a function of magnitude, that
    were potential targets (open histogram) and those that were
    observed (closed histogram). The top row gives these details for
    the galaxies in the clusters A1139, A3558 and A0930 (from left to
    right) while the bottom row gives these details for the galaxies
    in their outskirts.}
  \label{completeness}
\end{figure*}

In compiling the list of potential targets, no consideration was given
to morphological type, as that information was not
available. Classification of galaxies in the final sample was based on
inspection of their spectra (i.e.\ by spectral type, not morphological
type).

To ensure that our potential targets were cluster members (or in the
case of the 6dF observations, members of structures associated with
the clusters) the samples were trimmed of galaxies outside a $\pm 3
\sigma_{\mathrm{clus}}$ slice in redshift space centred on the cluster
redshift, where $\sigma_{\mathrm{clus}}$ is the cluster velocity
dispersion. This was not done for the galaxies in the Coma field, as
they were all confirmed cluster members.

When deciding which targets to observe, the highest priority was given
to the brightest galaxies with correspondingly lower priorities to
fainter galaxies. This had the dual effect of biasing the sample
toward the very bright galaxies in the core of the clusters and
ensuring that good S/N was achieved for the greatest number of objects
in the sample. Histograms showing the number of potential targets and
the number of galaxies actually observed as a function of magnitude
are given in Figure \ref{completeness}, where the top row is for
cluster galaxies and the bottom row is for galaxies in the cluster
outskirts and their surrounds. Coma is not shown here since galaxies
form this cluster were selected in a different manner (see above). It
is evident from this figure that we were successful in targeting the
brightest galaxies in each cluster and in their outskirts.

Stars from the Lick/IDS catalogue \citep{worthey94a} were selected as
velocity templates for the 2dF (HR6159, HR6770, and HR7317) and 6dF
observations (HR2574, HR3145, and HR5888). Details of these six stars
are given in Table \ref{stan stars}. They were also used as Lick
standards along with the Coma cluster galaxies as they could be
compared to the published line-strength data of \citet{moore02}.

\begin{table}
\begin{minipage}{118mm}
\caption{The stars that were observed as stellar standards.}
\begin{tabular}{ccccc}
\hline
Name & R.A. & Dec. & $\mathrm{m_B}$ & Spectral type\\
\hline
HR2574 & 06 54 11.40 & $-12$ 02 19.1 & 5.54 & K4III3\\
HR3145 & 08 02 15.94 & $+02$ 20 04.5 & 5.67 & K2III1\\
HR5888 & 15 50 17.55 & $+02$ 11 47.4 & 6.24 & G8III5\\
HR6159 & 16 32 36.29 & $+11$ 29 16.9 & 6.33 & K4III1\\ 
HR6770 & 18 07 18.36 & $+08$ 44 01.9 & 5.60 & G8III2\\
HR7317 & 19 19 00.10 & $-15$ 32 11.7 & 7.49 & K4III8\\
\hline
\end{tabular}
\label{stan stars}
\medskip

Coordinates are given in J2000.0.
\end{minipage} 
\end{table}

\subsection{Observing method}\label{obs}

\begin{table*}
\begin{minipage}{118mm}
\caption{A summary of the instrumental set-up of 2dF and 6dF.}
\begin{tabular}{lcc}
\hline
& 2dF & 6dF\\
\hline
Telescope & 3.9 m AAT & 1.2 m UKST\\
FoV (diameter) & $2^\circ$ & $5.7^\circ$\\
Grating & 300B & 580V \\
& 1200B & \\
& 1200V & \\
Spectral range & 3650--8050 \AA & 3920--5600 \AA \\
& 4590--5730 \AA & \\
& 4580--5720 \AA & \\
Dispersion & 4.3 \AA $\perpix$ & 1.6 \AA $\perpix$ \\
& 1.1 \AA $\perpix$ & \\
& 1.1 \AA $\perpix$ & \\
Instrumental resolution (FWHM) & 8.9 \AA & 5.9 \AA \\
& 2.2 \AA & \\
& 2.2 \AA & \\
Number of object fibres & 400 & 150 \\
Number of guide fibres & 4 & 4 \\
Object fibre diameter & 140 $\muem$ ($2.1\arcsec$) &
100 $\muem$ ($6.7\arcsec$)\\
Positioner accuracy & $\sim 20 \muem$ ($\sim 0.3\arcsec$) &
$\sim 10 \muem$ ($\sim 0.7\arcsec$)\\
Detector & 2 $\times$ thinned Tektronix CCD & Marconi CCD47-10 \\
Size & $1024 \times 1024$ & $1032 \times 1056$ \\
Pixel size & 24 $\muem$  & 13 $\muem$ \\
Gain & 2.8 $e^-$ ADU$^{-1}$ & 0.6 $e^-$ ADU$^{-1}$ \\
Read-out noise & 5.2 $e^-$ & 2.8 $e^-$ \\
\hline
\end{tabular}
\label{low red specs}
\end{minipage}
\end{table*}

Galaxies in the four clusters and three standard stars (HR6159,
HR6770, and HR7317) were observed with 2dF during the nights of 19--21
April 2002 using the 300B and 1200V gratings. The 300B grating was
used to maximise the number of Lick indices that could be measured
while the 1200V grating was used to enable accurate redshift and
velocity dispersion measurements. There was insufficient time to
obtain spectra of the A1139 galaxies with the 1200V grating; these
were were subsequently obtained on the night 5 January 2003 using the
1200B grating. The mean S/N per {\AA}ngstr{\"o}m (calculated at the
\mgb\ index) was 25.6 for the 300B observations; 36.2 for the 1200V
observations; and 27.5 for the 1200B observations. Multiple
observations were made of each galaxy with exposure times of
1200--1800 s and total exposure times ranged from 1--2 hours for the
300B grating and 2--6 hours for the 1200V/1200B. Exposure times were
3--10 s for the stellar standards, which were taken at the end of
every night. Each block of exposures was preceded by an arc exposure
and a flat-field exposure, and followed by another arc
exposure. Offset-sky exposures of 300 s were taken to aid in sky
subtraction. The instrumental set-up of 2dF is detailed in Table
\ref{low red specs}.

Galaxies in the outskirts of the four clusters and the structures
surrounding them, and three standard stars (HR2574, HR3145, and
HR5888) were observed with 6dF using the 580V grating during the
nights of 6--8 March 2003. The mean S/N per {\AA}ngstr{\"o}m
(calculated at the \mgb\ index) of the observations was 17.8. Multiple
observations of the galaxies were taken with exposure times of 1200 s,
while exposures of 5--10 s were taken of the stellar standards at the
beginning and end of every night. The total exposure times for each
cluster ranged from 2--4 hours. Each block of exposures was preceded
by an arc lamp exposure and a quartz lamp exposure for mapping the
positions of the spectra on the CCD. The 6dF sample contained galaxies
in common with the 2dF sample, but mostly consisted of those in the
outer regions of each cluster. The instrumental set-up of 6dF is
detailed in Table \ref{low red specs}.

\subsection{Basic reductions}\label{reductions}
 
The data were reduced using dedicated reduction pipelines, 2dfdr
\citep{colless01} and 6dfdr \citep{jones04}. Basic reduction steps
carried out by these programs include bias subtraction, spectrum
extraction, flat-fielding, wavelength calibration, fibre throughput
determination and correction, and sky subtraction.

The 6dF images were flat-fielded using normalised flat-field images
but the 2dF images were not flat-fielded because flexure in the
spectrographs, which were mounted at prime focus, means that the flat
field spectra are extracted from slightly different pixels to those
being calibrated. With both instruments the bias level was determined
by taking the median of the over-scan region.

The RMS error of the wavelength calibration for the 2dF spectra was
$0.30\pm 0.08$ \AA\ (300B) and $0.09\pm 0.02$ \AA\ (1200B/1200V). For
the 6dF spectra it was $0.05\pm 0.02$ \AA. In all cases, the accuracy
of the fit was better than one-tenth of a pixel.

The method used to calculate the fibre throughput varied. For the 300B
2dF spectra and the 6dF spectra the relative throughput of each fibre
was calculated from the flux in the 5577 \AA\ sky-line while for the
1200V and 1200B 2dF spectra, the offset-sky method was used. In all
cases, a mean sky spectrum was calculated from the spectra obtained
with the sky fibres, which was then subtracted from each spectrum
after being scaled to the appropriate throughput value.

For 2dF, the sky subtraction accuracy, measured by the residual light
in the sky fibres after sky subtraction, is 3.2\% (300B), 5.7\%
(1200V), and 5.2\% (1200B). For 6dF it is 3.0\%.

\section{Redshifts and Velocity Dispersions}\label{red and sig}

\subsection{Redshifts}\label{redshifts}

Redshifts were measured using the program runz \citep{colless01},
which cross-correlates the galaxy's spectrum with numerous spectral
templates. The IRAF task rvcorrect was then used to correct the
observed redshifts to heliocentric redshifts.

\begin{figure*}
\includegraphics[width=0.9\textwidth]{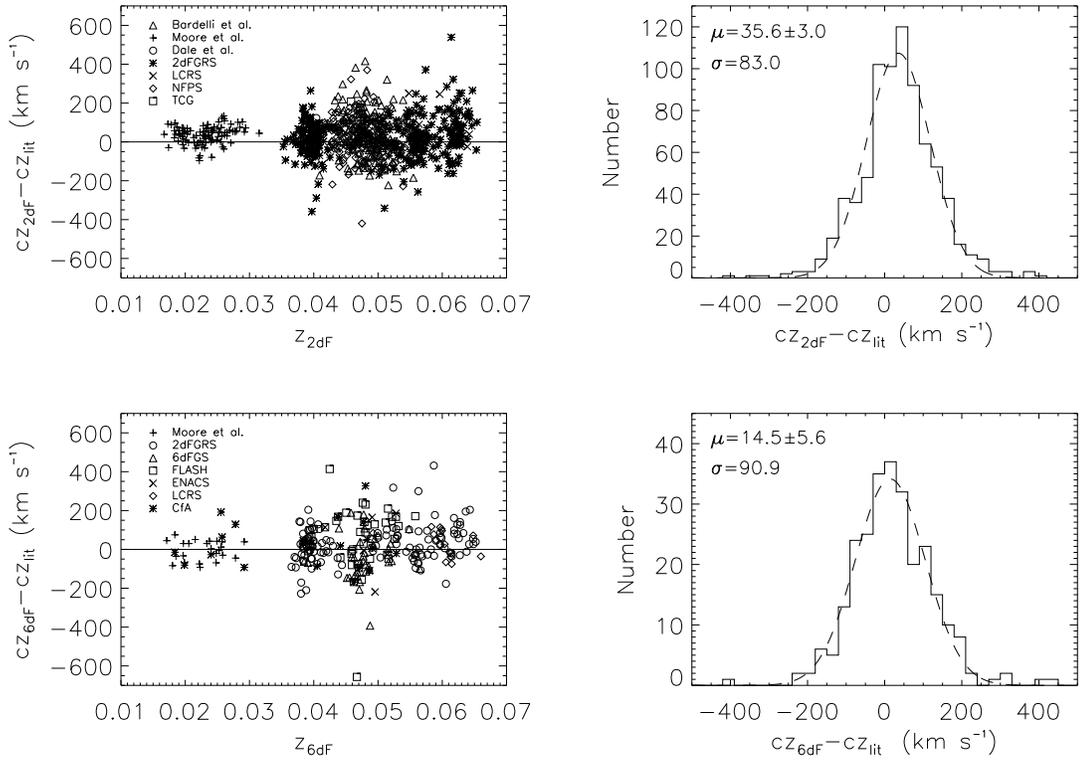}
\caption{Top left: Comparison of the redshifts measured for galaxies
  observed with 2dF to those from the literature. The keys to the
  meaning of each symbol are provided in the top left corner
  (references are provided in the text). Top right: Histogram of the
  redshift differences between galaxies observed with 2dF and those in
  the literature. The dashed curve is the best-fitting Gaussian; the
  mean and standard deviation are given in the top left corner. The
  bottom two panels are the same as the top two but are for galaxies
  observed with 6dF.}
\label{comp_lowz_z}
\end{figure*}

The top two panels in Figure \ref{comp_lowz_z} show a comparison of
the heliocentric redshifts measured for galaxies observed with 2dF to
redshifts published in the literature and a histogram of the redshift
differences. The dashed curve in the right panel is the best-fitting
Gaussian for the differences with the mean and standard deviation
given in the top left corner. 778 redshifts were obtained from the
literature: 293 of these came from 2dFGRS \citep{colless01}, 24 from
LCRS \citep{shectman96}, 10 from \citet{dale97}, 141 from
\citet{bardelli94, bardelli98}, 10 from \citet*[TCG;][]{teague90}, 202
from NFPS \citep{smith04}, and 98 from \citet{moore02}. The agreement
between this project and the literature is generally good. However, as
can be seen from the histogram, there is a statistically significant
offset. This offset is of the order of the precision of our redshift
measurements and translates to a spectral shift of $\sim 0.5$
\AA. Such a shift will have a minimal impact on the measurement of the
Lick indices that have central bandpasses on average 35 \AA\ in
width. Therefore, the redshifts were not corrected for this offset.

The bottom two panels of Figure \ref{comp_lowz_z} show the equivalent
plots for the 6dF data. 266 redshifts of galaxies in and around our
four clusters were obtained from the literature: 23 of these redshifts
came from \citet{moore02}, 20 from the CfA redshift survey
\citep*[CfA;][]{huchra99}, 123 from 2dFGRS, 17 from LCRS, 36 from
6dFGS \citep{jones04}, 41 from the FLASH redshift survey
\citep{kaldare03}, and 6 from ENACS \citep{katgert98}. There is no
statistically significant offset between the 6dF data and the
literature. For both the 2dF and 6dF observations the dispersion of
the differences between the measured redshifts and the literature is
$\sim 85 \kmpers$.

\begin{figure*}
  \includegraphics[width=0.9\textwidth]{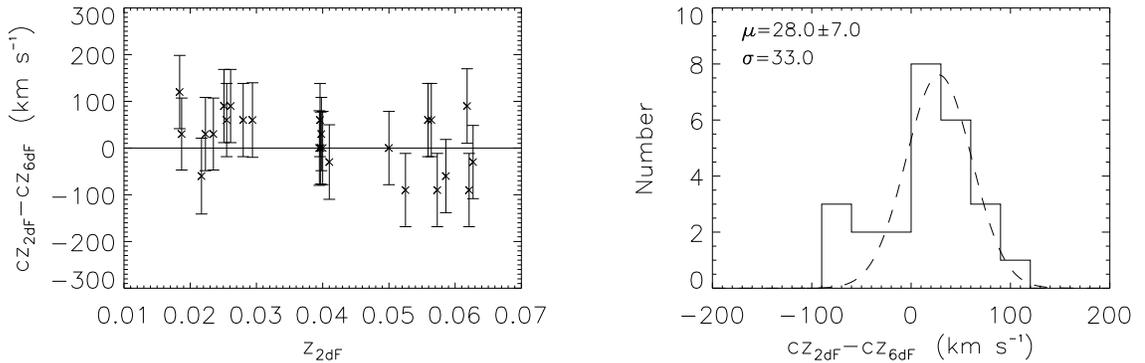}
  \caption{Left: Comparison of pairs of redshifts measured for
    galaxies with 2dF and 6dF. Right: Histogram of the redshift
    differences.}
  \label{comp_2df_6df_z}
\end{figure*}

25 galaxies were observed with both 2dF and 6dF and have two reliable
redshift estimates and in Figure \ref{comp_2df_6df_z} we compare these
estimates. There is a statistically significant offset between the two
datasets but again it is less than the precision of our redshift
measurements. The dispersion of the differences between the two
redshift estimates is $33 \kmpers$, which is also of the order of our
redshift precision.

\subsection{Velocity dispersions}\label{dispersions}

The IRAF task fxcor was used to measure the galaxy velocity
dispersions. The stars HR6159, HR6770, and HR7317 were used as
velocity templates for cross-correlating with the galaxies observed
with 2dF, while HR2574, HR3145, and HR5888 were used for the galaxies
observed with 6dF (see Table \ref{stan stars} for details). Each of
these stars were observed twice, providing six velocity templates for
cross-correlating.

Before the velocity dispersions were measured, each spectrum was
processed to ensure as close a match as possible with the template
against which it was to be cross-correlated. Each galaxy was
de-redshifted using the IRAF task dopcor and the redshift provided by
runz. The template spectrum was then re-binned so as to match the
dispersion of this de-redshifted spectrum. Both spectra were trimmed
so that they contained only the wavelength range in common and
broadened to the lowest resolution exhibited by either spectrum, since
both 2dF and 6dF have instrumental resolutions that are fibre and
wavelength dependent.

\begin{figure*}
\includegraphics[width=0.9\textwidth]{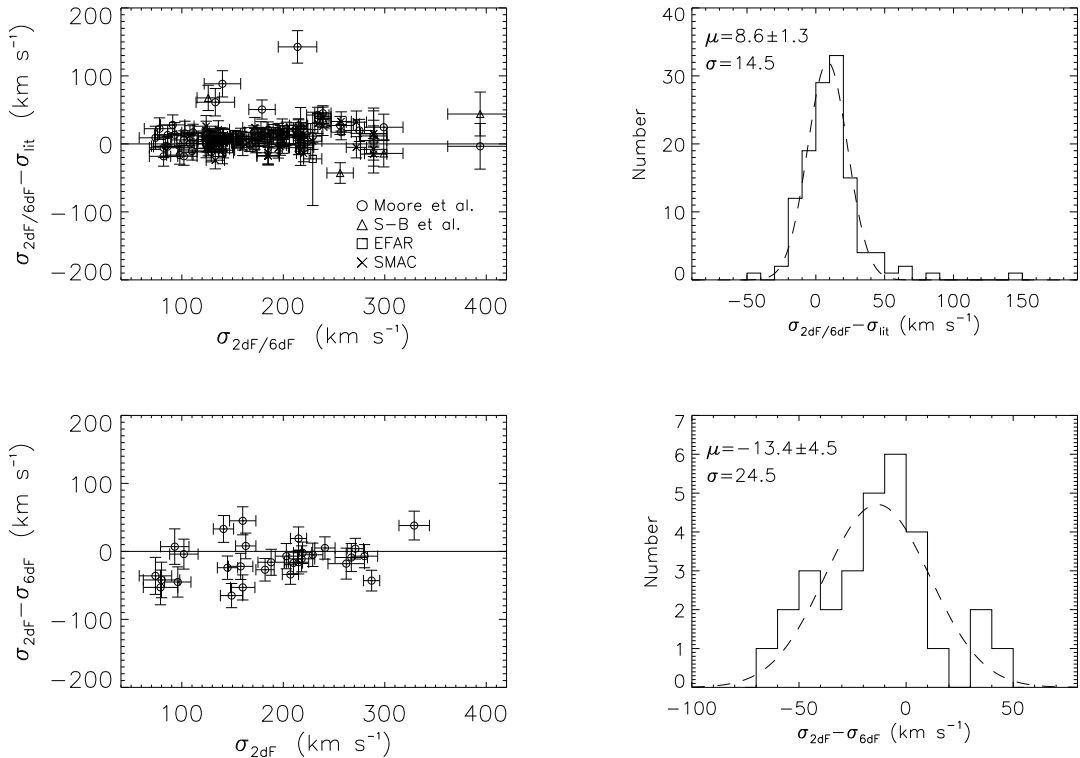}
\caption{Top left: Comparison of the velocity dispersions of the
  galaxies measured with 2dF and 6dF to those published in
  literature. The solid line has a slope of unity and the dashed line
  is our fit to the data. The key to the symbols are shown in the top
  left corner and references are given in the text. Top right:
  Histogram of velocity dispersion differences. The dashed curve is
  the best-fitting Gaussian for the data, the mean and standard deviation
  are given in the top left corner. The bottom two panels are the same
  as the top two but compare estimates of the velocity dispersion for
  galaxies observed with both 2dF and 6dF.}
\label{comp_sig}
\end{figure*}

The above procedure was carried out on each galaxy/template
combination and a velocity dispersion measured. The final velocity
dispersion was taken as the mean of those derived from the six
templates, weighted by the cross-correlation R-value. The top two
panels in Figure \ref{comp_sig} compare the measured velocity
dispersions to those published in the literature. There are 124
velocity dispersions against which we compare our estimates. Of these
literature velocity dispersions, 73 are taken from \citet{moore02}, 8
from \citet{sanchez06}, 5 from EFAR \citep{wegner99}, and 38 from SMAC
\citep{smith00}. At low velocity dispersions the agreement is good,
but at $\sigma> 160 \kmpers$ our estimates seem to be greater than
those in the literature. This offset is mainly caused by the
\citeauthor{moore02}\ data. At $\sigma > 180 \kmpers$ there is an
offset between this dataset and ours, with our estimates being on
average $\sim 17.6 \kmpers$ larger. A similar offset of $\sim 17.4
\kmpers$ to the \citeauthor{moore02}\ data was found by
\citet{sanchez06}. The right panel shows a histogram of the velocity
dispersion differences and the best-fitting Gaussian for the data
(dashed curve). The mean and the standard deviation, which are given
in the top left corner, indicate that there is a small but significant
offset of $\sim 8 \kmpers$ and that the dispersion of the differences
between the estimates is $\sim 15 \kmpers$.

The bottom two panels in Figure \ref{comp_sig} are identical to the
top two but compare the velocity dispersions measured for galaxies
observed with both 2dF and 6dF.  We see from the left panel that there
is large scatter at lower velocity dispersions possibly due to the
lower luminosity, and hence lower S/N, of these galaxies. Another
possible reason for the scatter could be the aperture corrections (see
Section \ref{ap corr}). While the velocity dispersions have been
corrected for the differing aperture sizes, it is possible that at
lower velocity dispersions the 6dF fibres subtend a linear distance
larger than the diameter of the galaxy resulting in an over-correction
to the velocity dispersion, in the sense that the velocity dispersion
is larger than it should be.  The mean of the best-fitting Gaussian for
the velocity dispersion errors shows no significant offset between the
two datasets, however the dispersion of the differences is $\sim
25 \kmpers$.

\subsection{Redshift and velocity dispersion errors}\label{errors}

The program runz does not provide any estimate of the error associated
with a measured redshift, nor does fxcor provide an estimate of the
velocity dispersion error. Following \citet{wegner99}, we estimate
redshift and velocity dispersion errors from detailed Monte Carlo
simulations of the measurement process. The procedure used here, which
is the same for 2dF and 6dF, determines both the redshift and velocity
dispersion errors simultaneously as follows.

For each of the six 2dF stellar templates (there were two spectra for
each standard) we created a set of simulated spectra with S/N ranging
from 10 to 90 in steps of 10. Each spectrum in the set was then
Gaussian broadened to give spectra with dispersions from $60 \kmpers$
to $400 \kmpers$ in steps of $20 \kmpers$. Ten realisations of each of
these 972 spectra were then generated with Poisson noise. At a given
S/N and Gaussian broadening, each simulated spectrum was
cross-correlated with the original spectra from the other two stellar
templates giving us 240 redshift and velocity dispersion
estimates. The procedures used to prepare the simulated spectra and
the templates and the settings used in fxcor were identical to those
used in the actual measuring of the redshifts and velocity
dispersions. The only difference is that in the actual measurements
there was a spectral mismatch between the galaxy spectra and the
stellar template. However, at the S/N typical of our data this effect
is dominated by the Poisson noise.

Using the above method, the mean redshift error is $80 \kmpers$ and
the mean velocity dispersion error is 8\%. The measured redshifts,
velocity dispersions, and the estimated errors are given in Table
\ref{table_a1} of Appendix \ref{gal cats}.

\section{Line-strength Indices}\label{indices}

Detailed discussion of the Lick system can be found in any of the
references given in Section \ref{intro}. Some of the original index
definitions have been modified and it is the version of the index
definitions published in \citet{trager98}\ that is used in this
study. We also use the indices $\hbetag$, [OIII]$_1$ and [OIII]$_2$ as
defined in \citet{gonzalez93}.

To make valid comparisons with other datasets, it is essential that
the data are correctly transformed to the Lick system. The Lick system
and its associated models \citep[e.g.][]{worthey94b, vazdekis96,
thomas03a} are calibrated at an instrumental resolution of $\sim
9$ \AA\ FWHM for stellar populations with zero velocity dispersion.

Thus the spectra must be broadened to match the resolution of the Lick
system before line-strengths are measured and then a correction must
be applied to compensate for the change in index strength caused by
the broadening resulting from the velocity dispersion of the stellar
population. Generally, the effect of velocity dispersion broadening is
to weaken the line-strength, as the absorption line is broadened past
the limits of the central bandpass, although the result is actually a
complicated combination of effects relating to the movement in the
levels of the flanking pseudo-continuum as well as the change in width
of the spectral feature.

The final step in transforming the data to the Lick system is to apply
a correction to an index if it is offset from literature data
calibrated to the Lick system. These systematic offsets can be caused,
for example, by continuum shape differences. The index Mg$_2$ has an
offset that is of the order of 0.02 mag and is caused by the fact
that the original Lick spectra are not flux-calibrated. The procedures
used here to transform the data closely follow those outlined in a
number of papers \citep[e.g.][]{gonzalez93, fisher95, worthey97,
trager98, kuntschner00}.

The Lick spectra cover the wavelength range 4000--6000 \AA\ and have a
mean instrumental FWHM of $\sim 9$ \AA, increasing at both ends. The
instrumental FWHM of the 2dF and 6dF spectra show similar trends with
wavelength as well as a fibre dependence. Thus, it is necessary to
perform a wavelength-dependent as well as fibre-dependent
broadening. The low-resolution 2dF data were of a comparable or
slightly lower resolution than the Lick spectral resolution and as
such they were not broadened.

Line-strengths were measured using a program called indexf
\citep{cardiel98}. indexf is a highly versatile program that allows
for the measurement of line-strength indices as well as the estimation
of errors and S/N per \AA\ in wavelength-calibrated FITS
files.\footnote{Note: the radial velocity required by indexf must be
calculated using the relativistic formula and not just c$z$.} Positive
values correspond to lines that are in absorption while negative
values correspond to those in emission.

None of the spectra obtained in this project was flux calibrated, so
to minimise the effects of continuum shape differences, all spectra
had their continuum divided out. A low-order polynomial was fit to the
continuum of each spectrum and then this fit was divided out.

For the 2dF galaxies all 21 Lick indices, $\hbetag$, [OIII]$_1$, and
[OIII]$_2$ were measured. For the 6dF galaxies 16 Lick indices (from
CN$_1$ to Fe5406), $\hbetag$, [OIII]$_1$, and [OIII]$_2$ were
measured. However, because these indices need to be corrected for
offsets to the Lick system (as mentioned above) and for aperture
effects, and because these corrections are not available for all
indices, the actual usable indices for the 2dF/6dF galaxies are:
C$_2$4668, $\hbeta$, $\hbetag$, Fe5015, Mg$_1$, Mg$_2$, \mgb, Fe5270,
Fe5335 and Fe5406. Aperture corrections are not available for
[OIII]$_1$, [OIII]$_2$ but these indices were used only to identify
emission-line galaxies.

\subsection{Line indices corrections}

\subsubsection{Velocity dispersion corrections}\label{sigma corrections}

The Lick system is calibrated for stellar populations with zero
velocity dispersion. Correction factors must therefore be calculated
and applied once the line-strengths have been measured. The effect of
velocity broadening is usually to decrease the strength of the index.

The correction factors were determined by empirically measuring the
effect of an increasing velocity dispersion on the measured
line-strengths. To do this, the standard stars were broadened out to
the Lick resolution, using the method described above, and then
broadened further in steps of $50 \kmpers$ out to a velocity
dispersion of $400 \kmpers$. At each step the line-strength of each
index was measured. A third-order fit was then made for each index to
create a relation for which, given a velocity dispersion, a correction
factor can be determined. This correction factor is multiplicative for
atomic indices and additive for molecular indices.

\subsubsection{Aperture corrections}\label{ap corr}

\begin{figure*}
\includegraphics[width=0.9\textwidth]{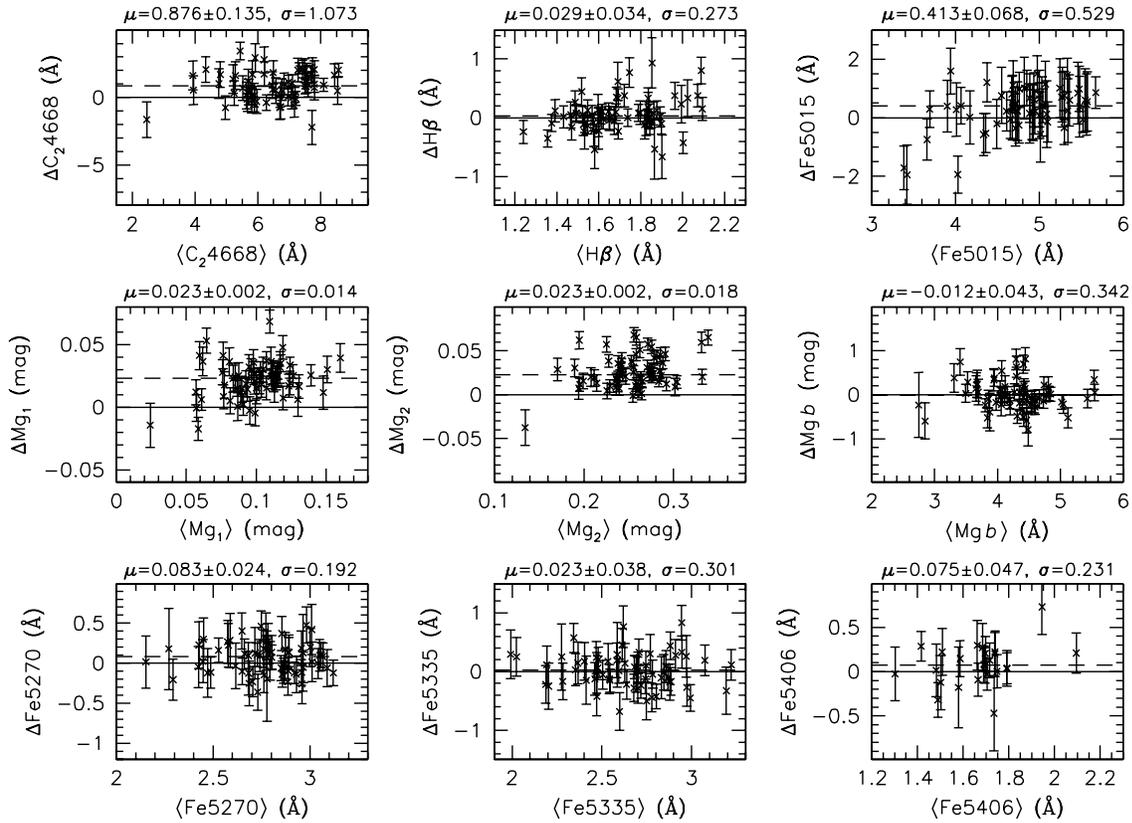}
\caption{Offsets between the line-strengths of Coma galaxies measured
with 2dF and those found in \citet{moore02}. The solid line is an
offset of zero and the dashed line is the measured offset. The value
of the offset and its error, and the scatter about the offset, are
given at the top of each panel.}
\label{moore 2df offsets}
\end{figure*}

It has been known for a long time that elliptical galaxies exhibit
line-strength gradients \citep[e.g.][]{gorgas87}. Thus, corrections
must be applied to our measured line-strengths to account for the
different physical sizes subtended by fibres of different diameters on
galaxies at different redshifts.

Therefore, all line-strengths and velocity dispersions were corrected
to the smallest physical size subtended by any of the fibre/redshift
combinations used in this project $\sim 1.0$ kpc, i.e.\ 2dF fibres
($2.1\arcsec$ in diameter) at the mean redshift of the Coma cluster
($z=0.023$). The average line-strength gradients found in
\citet{kuntschner02} were used to perform these aperture
corrections. Gradients are available for only 9 indices: C$_2$4668,
$\hbeta$, Fe5015, Mg$_1$, Mg$_2$, \mgb, Fe5270, Fe5335, Fe5406. So
subsequent analysis will include only these indices. We also include
$\hbetag$ in the analysis because we can use the gradient found for
$\hbeta$, which we expect to be the same, to perform the aperture
corrections. We note that the gradient for H$\gamma_\mathrm{A}$
(0.034) is almost identical that of H$\gamma_\mathrm{F}$ (0.033). In
any case, no gradient was found for $\hbeta$ and so no correction was
applied to either $\hbeta$ or $\hbetag$. The form of the correction is
identical to that found in \citet{jorgensen97}. The logarithmic
velocity dispersion aperture corrections are applied in a manner
identical to the line-strength corrections with $\alpha=-0.04$
\citep*{jorgensen95}.

\subsection{Calibrating the data}\label{cali}

The final step in putting the data on to the Lick system is to compare
it with published data already calibrated to the Lick system, taking
into account the aperture effects mentioned above, and checking for
any offsets. For this purpose, Coma cluster galaxies were observed
with both 2dF and 6dF.

\begin{figure*}
\includegraphics[width=0.9\textwidth]{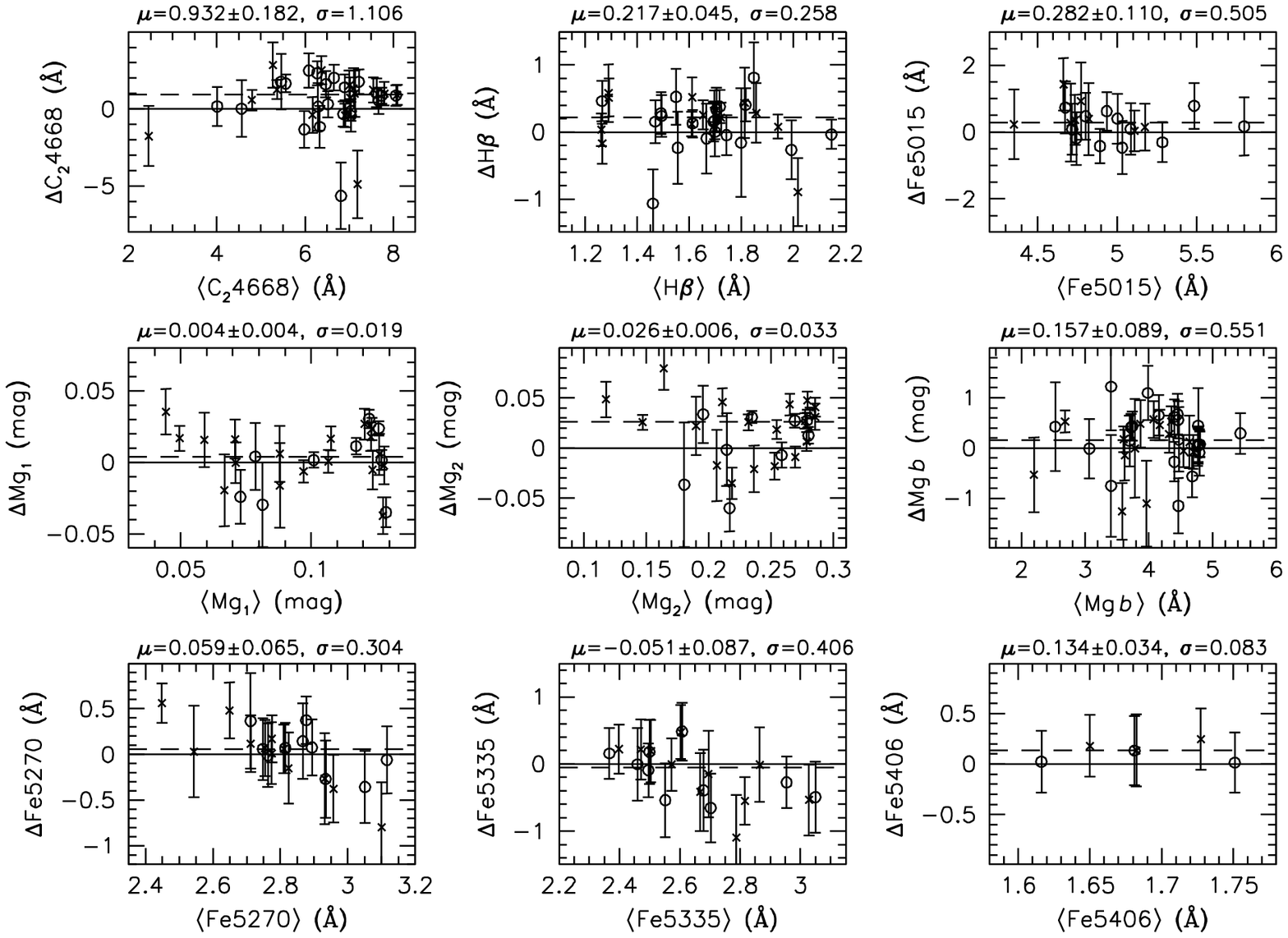}
\caption{Offsets between the line-strengths of Coma galaxies measured
with 6dF and those found in \citeauthor{moore02}\ (crosses), and the
fully calibrated 2dF data (open circles). The solid line is an offset
of zero and the dashed line is the measured offset. The value of the
offset and its error, and the scatter about the offset, are given at
the top of each panel.}
\label{moore 6df offsets}
\end{figure*}

Plots of the offsets between our data and that in the literature for
all indices for which aperture corrections were applied are shown in
Figures \ref{moore 2df offsets} and \ref{moore 6df offsets}. In each
panel the solid line represents zero offset while the dashed line
represents the mean offset that needs to be added to our data to bring
them into agreement with the literature values. Indices with an offset
significant at greater than the $3\sigma$ were corrected.

For the 2dF data (Figure \ref{moore 2df offsets}), we calibrate our
data by comparing to that of \citet{moore02}. Significant offsets were
found for C$_24668$, Fe5015, Mg$_1$, Mg$_2$, and Fe5270. For the 6dF
data (Figure \ref{moore 6df offsets}), we compare our data to the Coma
cluster galaxies data of \citeauthor{moore02}\ as well as our own
fully calibrated 2dF data. Including the 2dF data allows us to
increase the number of galaxies in the comparison and to check for
offsets between the 2dF and 6dF data before combining the data of
galaxies in common. Significant offsets were found for C$_24668$,
$\hbeta$, $\hbetag$, Mg$_2$, and Fe5406.

With the exception of C$_24668$ and Fe5015 (in both the 2dF and 6dF
data) the offsets found are all physically small, being on average
only 0.024 mag for molecular indices and 0.143 \AA\ for atomic
indices. The offset for Fe5406 (6dF data) was determined from just six
data points but since all showed an offset in the same direction this
offset was still applied. The offsets for $\hbetag$ were, within the
errors, consistent with those of $\hbeta$ and so we only show the
$\hbeta$ offsets.

\section{Spectral classification}\label{detect sf}

As this project is primarily focused on the old stellar populations of
early-type galaxies, all galaxies that exhibited signs of star
formation were excluded from our sample of early-type
galaxies. H$\alpha$ and [OIII]$\lambda$5007 \AA\ emission were used
to determine whether a galaxy was star-forming or not. Only the 2dF
300B observations had enough wavelength coverage to measure the
strength of H$\alpha$, so for the 6dF observations we relied on
[OIII]$\lambda$5007 \AA\ alone.

H$\alpha$ emission was determined for galaxies observed with 2dF using
the de-blending function in the IRAF task splot. Fitting a Gaussian to
these data reveals no offset from zero emission and that our
measurement errors are $\sim 1.9$ \AA. Galaxies that showed H$\alpha$
emission more than $2\sigma$ from zero, which equates to H$\alpha$
emission less than $-3.8$ \AA\ (lines in emission have negative
line-strengths in the Lick system), were removed from our sample.

Even with these galaxies removed there were galaxies remaining that
showed signs of significant [OIII]$\lambda$5007 \AA\ emission, as
measured by the index [OIII]$_2$. Statistically, $\hbeta$ emission has
been found to correlate with [OIII]$\lambda$5007 \AA\
\citep{gonzalez93}. Galaxies that show signs of
[OIII]$\lambda$5007 \AA\ emission will therefore have weaker $\hbeta$
line-strengths (due to nebular $\hbeta$ emission filling in the
stellar $\hbeta$ absorption) and thus incorrectly older ages. The
usual procedure in these cases is to correct the $\hbeta$
line-strength by subtracting from it a fraction of the [OIII]$_2$
line-strength. Two fractions have been suggested and used in the
literature: 0.7 \citep{gonzalez93} and 0.6
\citep{trager00a,trager00b}. Recently, the use of
[OIII]$\lambda$5007 \AA\ emission to correct for $\hbeta$ emission,
even in a statistical sense, has been questioned \citep{nelan05}. The
ratio of emission from these two lines spans a large range and
galaxies with little to no $\hbeta$ emission can show significant
[OIII]$\lambda$5007 \AA\ emission and vice versa.

Therefore, to be certain that the early-type galaxy sample is free
from star-forming galaxies, we also exclude any galaxy that shows
significant [OIII]$\lambda$5007 \AA\ emission. Galaxies with an
[OIII]$_2$ index less than $-0.4$ \AA\ were deemed to be star-forming
and were therefore excluded from the early-type galaxy sample. This
represented a 2$\sigma$ cut for the 2dF data and a 1$\sigma$ cut for
the 6dF data.

The fully-calibrated line-strengths and errors are given in Table
\ref{table_a2} of Appendix \ref{gal cats}.

\section{Stellar Population Parameters}\label{parameters}

The stellar population parameters age, [Z/H], and $\afe$ were
estimated using the $\chi^2$ method of \citet{proctor04} and comparing
our measured line-strengths to the models of \citet{thomas03a}. These
models are relatively coarse, predicting line-strengths for stellar
populations with ages between 1 Gyr and 15 Gyr, [Z/H] between $-2.25$
dex and 0.67 dex, and $\afe$ between 0.0 dex and 0.5 dex. The
predictions are made at 1 Gyr intervals in the age, but at a small
number of varying intervals in [Z/H] and $\afe$. To more accurately
determine the stellar population parameters, we interpolate the models
to a fine grid providing 378000 individual models, spanning [Z/H]$\sim
-1.19-0.59$ (in steps of 0.02 dex), age$\sim 1-15$ Gyr
(logarithmically, in steps of 0.02 dex), and [$\alpha$/Fe]$\sim
-0.19-0.50$ (in steps of 0.01 dex).

For each galaxy, the stellar population parameters corresponding to
the set of predicted line-strengths that produces the smallest
$\chi^2$ when compared with the observed line-strengths are adopted as
the values for that galaxy. As in \citeauthor{proctor04}, indices that
differed from the accepted fit by more than three times the RMS
scatter about it were rejected. The indices Mg$_1$ and Mg$_2$ were
found to be problematic and so were excluded from the fitting
procedure. Both these indices are very broad and possibly were not
calibrated correctly due to our dividing out of the continuum before
measuring the line-strengths. Besides, very few galaxies had measured
line-strengths for these indices as they were at the limit of our
wavelength coverage for 6dF and often were affected by sky-lines. This
left a maximum of seven indices with which to perform the fits:
C$_2$4668, $\hbeta$, Fe5015, \mgb, Fe5270, Fe5355 and Fe5406. A fit
was only accepted if $\hbeta$ was used and at least two other indices.

\begin{figure*}
\includegraphics[width=0.9\textwidth]{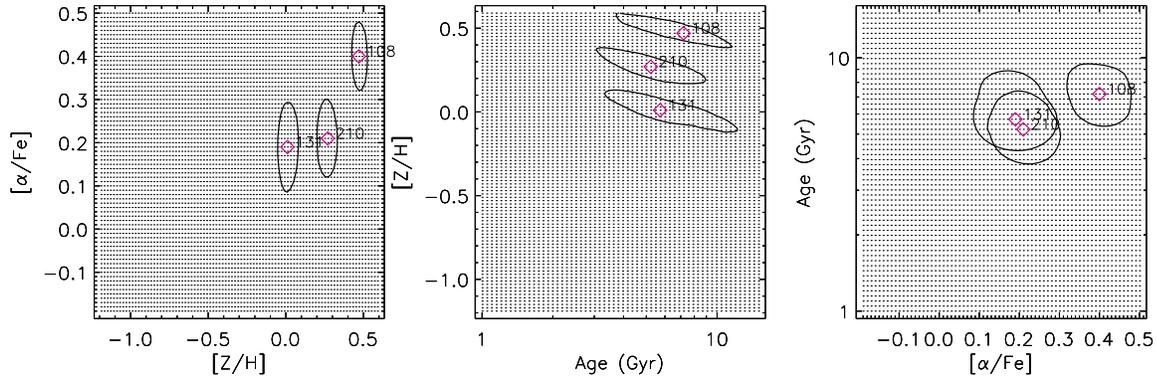}
\caption{Examples of the confidence limits for the parameter
  estimates. All three galaxies have roughly mean errors on their
  indices but span our range in velocity dispersion. One having a low
  velocity dispersion of $84 \kmpers$ (131), one with an average
  velocity dispersion of $171 \kmpers$ (210) and the last with a high
  velocity dispersion of $299 \kmpers$ (108). The grid of dots in
  each plot represent the models of \citet{thomas03a} interpolated as
  described in the text.}
\label{param errors}
\end{figure*} 

\begin{figure*}
\includegraphics[width=0.9\textwidth]{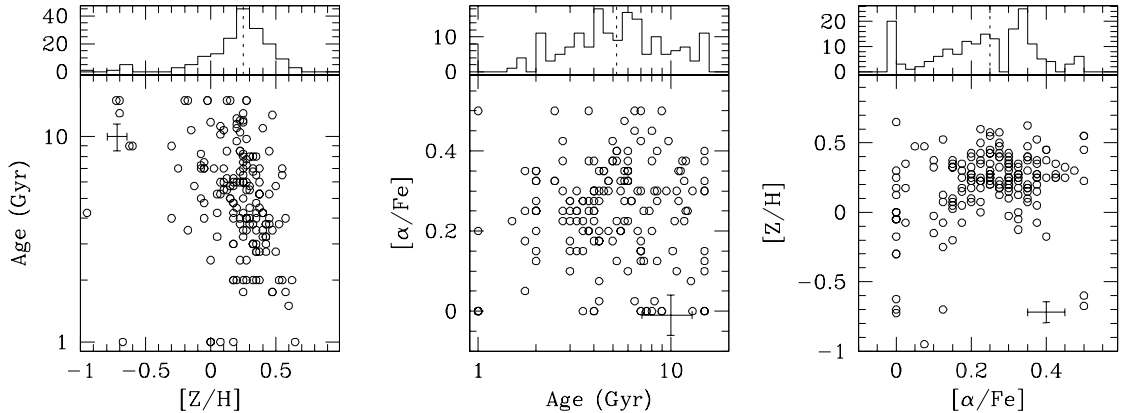}
\caption{The distributions of stellar population parameters for all
  galaxies in the sample. Marginal distributions are shown for each
  parameter, and median errors are given in each panel; the dotted
  lines show the median values of each parameter.}
\label{gal params all}
\end{figure*} 

Errors on the parameters were estimated by using the constant $\chi^2$
boundaries as confidence limits \citep[see][]{press92}. Due to the
irregular shape of the grids formed by the stellar population
parameters in index space these errors change with position on the
grid. Additionally the spacing between lines of constant age reduces
when moving to older ages. Therefore the quoted errors for the age
estimates are two-sided, but those for the other two parameters are
simply the largest of the two errors. As the error estimates cannot be
larger than the distance to the edge of the grid this means for some
galaxies the positive age errors will be lower limits. Examples of the
confidence limits for three galaxies are shown in Figure \ref{param
  errors}. These galaxies all had roughly mean errors on their indices
but were chosen to span our range of velocity dispersions, a low
velocity dispersion of $84 \kmpers$ (131), one with an average
velocity dispersion of $171 \kmpers$ (210) and the last with a high
velocity dispersion of $299 \kmpers$ (108).

The distributions of stellar population parameters for all galaxies in
the cluster sample are shown in Figure \ref{gal params all}. Marginal
distributions are given for each parameter and the dashed line in each
denotes the median value. The median [Z/H] is 0.25 dex, the median age
6.3 Gyr, and the median $\afe$ is 0.27 dex. Looking at the
distribution we see that that they are all roughly Gaussian and that
cluster early-type galaxies have mostly super-solar [Z/H], a large
spread in age, and $\afe$ that mostly falls between 0.2 dex and 0.4
dex. The interpretation of these results, and a discussion of the
assumptions made in determining these stellar population parameters,
will be the presented in a subsequent paper.

The stellar population parameter estimates and errors are given in
Table \ref{table_a3} of Appendix \ref{gal cats}.

\subsection{The effects of \boldmath $\afe$ on stellar population
  parameter estimates}\label{afe effects}

We now illustrate the problem of estimating ages and [Z/H] without
regard to the $\afe$ of the population. Using solar-based stellar
population models to derive age and [Z/H] estimates results in
over-estimated ages and under-estimated [Z/H] if an Fe index is used
as the [Z/H] indicator; the opposite will be true if an Mg index is
used. More generally, if the model used is based on a single $\afe$
then galaxies that have ratios larger than that will have
over-estimated ages and under-estimated [Z/H] if using an Fe index as
a [Z/H] indicator, and vice versa if an Mg index is used.

\begin{figure*}
\includegraphics[width=0.49\textwidth]{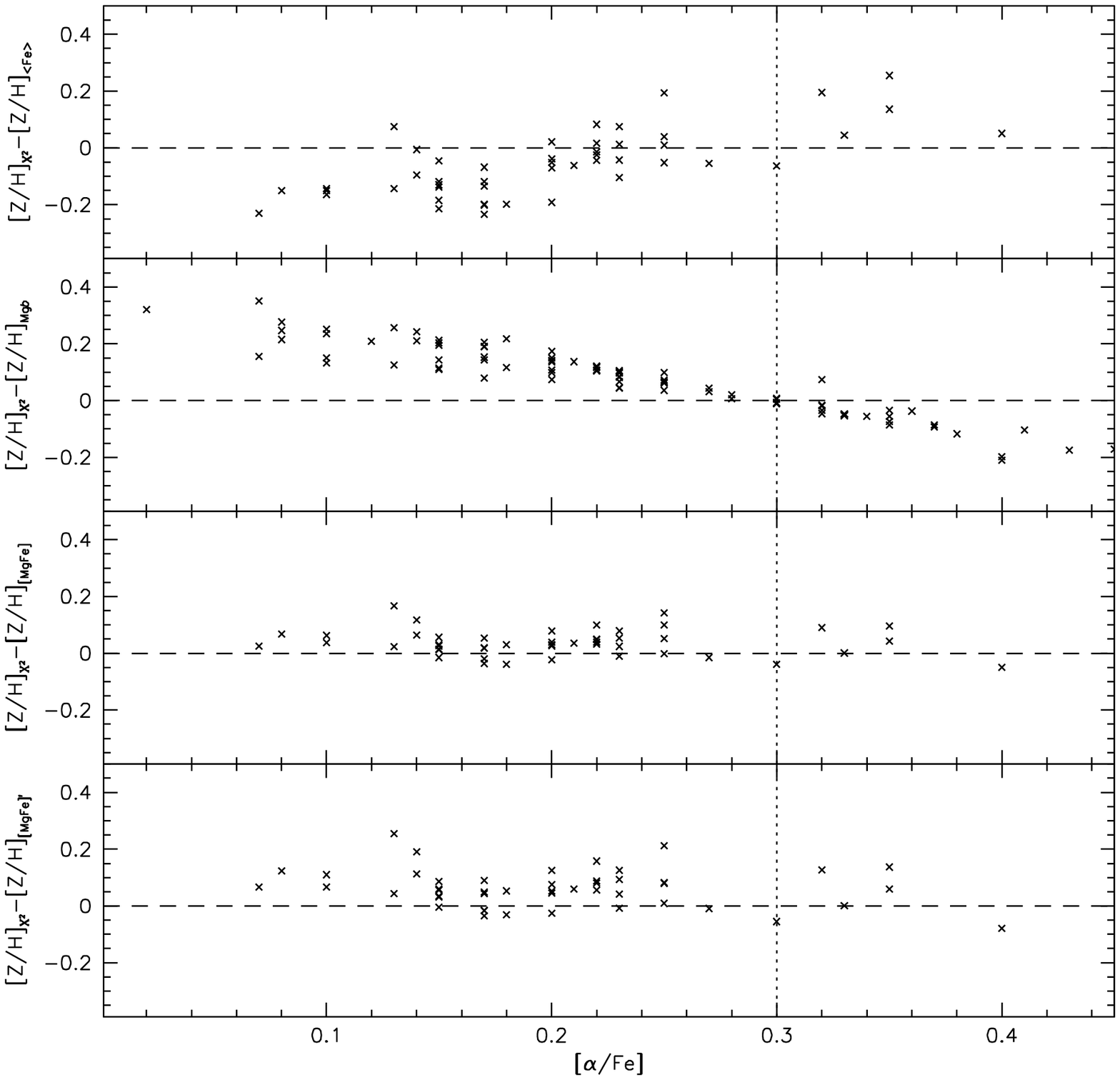}
\includegraphics[width=0.49\textwidth]{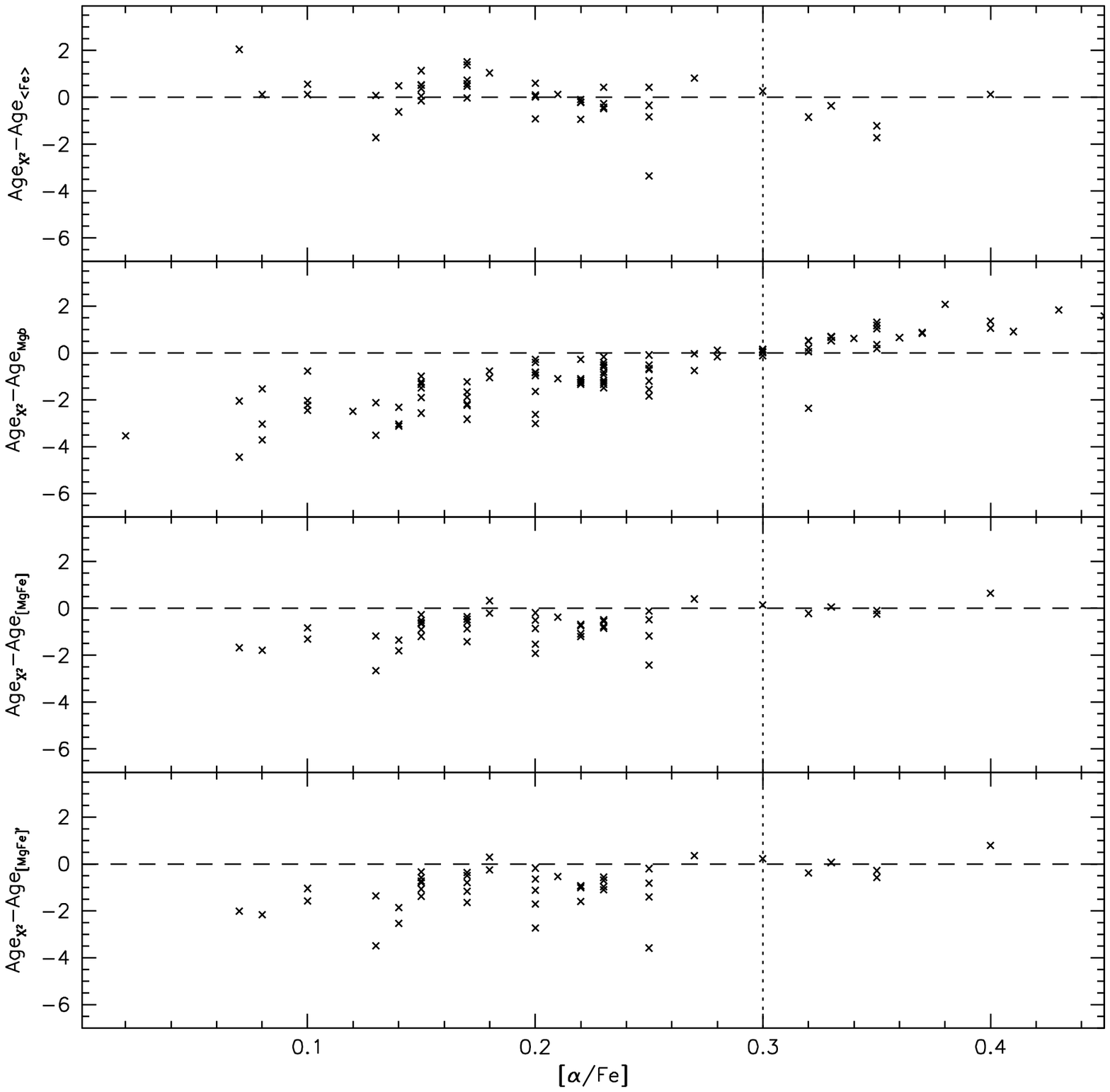}
\caption{The [Z/H] (left) and age (right) residuals as a function of
$\afe$. In each of the panels the parameter as estimated from the line
diagnostic diagram method is subtracted from the parameter as
estimated by the $\chi^2$ method. $\hbeta$, \mgb, and Fe5335 were
used in the $\chi^2$ method for all panels while the
[Z/H]-sensitive index used in the line diagnostic diagram method
was varied: $\langle \mathrm{Fe} \rangle$, \mgb, [MgFe], and
[MgFe]$^\prime$ (from top to bottom). In all cases $\hbeta$ was used
as the age-sensitive index. The dotted vertical line represents the
assumed $\afe$ for the line diagnostic diagram method.}
\label{alpha eff met}
\end{figure*}

We illustrate this by comparing our age and [Z/H] estimates from the
$\chi^2$ method, which has $\afe$ as a free parameter, to those
derived from line diagnostic diagrams where all galaxies are assumed
to have $\afe=0.3$ dex. Four different combinations of indices were
used; in all cases the age sensitive index was $\hbeta$ and the [Z/H]
indicator was one of \mgb, $\langle\mathrm{Fe} \rangle$, [MgFe]
\citep{gonzalez93}, or [MgFe]$^\prime$ \citep{thomas03a} -- these
latter two indices were designed to be insensitive to
$\afe$\footnote{$\langle\mathrm{Fe}\rangle
  =(\mathrm{Fe5270}+\mathrm{Fe5335}) /2$;\newline
  $[\mathrm{MgFe}]\equiv \sqrt{\mathrm{Mg\,\textit{b}}\times
    (\mathrm{Fe}5270+ \mathrm{Fe}5335)/ 2}$;\newline
  $[\mathrm{MgFe}]^\prime \equiv \sqrt{ \mathrm{Mg\,\textit{b}} \times
    (0.72 \times \mathrm{Fe}5270+ 0.28 \times \mathrm{Fe}5335)}$}. The
resulting differences between the [Z/H] estimated from the four line
diagnostic diagrams and the $\chi^2$ method as a function of $\afe$
are shown in the left panel of Figure \ref{alpha eff met}. The first
thing to note is that if the galaxy has an $\afe=0.3$ dex then the
estimates agree no matter which combination of indices are used. As
expected, at $\afe>0.3$ dex, $\langle\mathrm{Fe}\rangle$
under-estimates the [Z/H] while \mgb\ over-estimates it. The opposite
is true for each index at $\afe<0.3$ dex. For [MgFe] and
[MgFe]$^\prime$ there is no trend with $\afe$, but they consistently
under-estimate the [Z/H], with [MgFe]$^\prime$ slightly more so than
[MgFe].

The right panel of Figure \ref{alpha eff met} shows the equivalent
plots for the age estimates. Again we see that, at $\afe=0.3$ dex,
all methods agree, but at higher values $\langle\mathrm{Fe}\rangle$
over-estimates the age and \mgb\ under-estimates it, while the
opposite is true at lower values. [MgFe] and [MgFe]$^\prime$
consistently over-estimate the ages at $\afe<0.3$ dex
([MgFe]$^\prime$ slightly more so than [MgFe]) but seem to give
reasonable estimates at values greater than this. This exercise
emphasises the importance of including $\afe$ estimates when
estimating ages and [Z/H] of galaxies.

\section{Summary}\label{summary}

In summary, we have measured velocity dispersions, redshifts and
line-strengths for a total of 416 galaxies. These line-strengths were
carefully calibrated to the Lick system, and corrected to reflect the
strength that would have been obtained using 2dF on galaxies at the
redshift of the Coma cluster, i.e.\ we measured line-strengths of the
stellar populations in the inner $\sim 1$ kpc of each galaxy. During
the entire process, particular attention was paid to ensuring all
sources of errors were properly accounted for.

\begin{table}
\caption{The numbers of galaxies with measured line-strengths and
  estimated stellar population parameters within the various sub-samples.}
\begin{tabular}{lcc}
\hline
Sample & Line-strengths & Parameters\\
\hline
Cluster & 158 & 142\\
Cluster-outskirts & 87 & 77\\
Emission-line & 168 & ---\\
Unclassified & 3 & ---\\
\hline
\end{tabular}
\label{samples}
\end{table}

In a subsequent paper we will explore the variations in the stellar
populations of early-type galaxies with redshift and environment using
a number of sub-samples defined on our data set -- see Table
\ref{samples}. These samples are as follows. 158 are early-type
galaxies from Coma, A1139, A3558, or A930 observed with 2dF, 6dF, or
both, that are located within the Abell radius (i.e.\ at a projected
radial distance $\le2 h_{70}^{-1}$ Mpc). This is the cluster
sample. 87 are early-type galaxies from the outskirts of these
clusters at projected radial distances $>2 h_{70}^{-1}$ Mpc. This is
the cluster-outskirts sample. 168 galaxies were determined, based on
H$\alpha$ and/or [OIII]$\lambda$5007 \AA\ emission, to be
star-forming. This is the emission-line sample. The 3 remaining
galaxies lacked the information necessary to classify them as either
an early-type or emission-line galaxy. We include them in our
catalogue because they have measured line-strengths but do not include
them in any of the above samples.

Using the fully-calibrated absorption-line indices coupled with the
stellar population models of \citet{thomas03a}, we have estimated
ages, [Z/H], and $\afe$ for 219 galaxies. 142 of these galaxies are in
the cluster sample (8 from A930, 8 from A1139, 65 from Coma, and 61
from A3558), and 77 are in the cluster-outskirts sample. In the
process we have examined the affect on parameter estimates of
non-solar abundance ratios, by comparing various sets of estimates
where [$\alpha$/Fe] was fixed to a set of estimates where
[$\alpha$/Fe] was a free parameter. We find that the composite indices
[MgFe] and [MgFe]$^\prime$, although being insensitive to
[$\alpha$/Fe], over-estimate ages and under-estimate [Z/H]. The
results from this investigation indicate the importance of allowing
[$\alpha$/Fe] to be estimated along with [Z/H] and age.

In the second paper of this series we will utilise these data to
investigate the stellar populations of early-type galaxies in clusters
and their outskirts. Line-strength--$\sigma$ relations will be
examined and compared for both the cluster sample and the cluster
outskirts sample. Likewise, we will examine and compare the
parameter--$\sigma$ relations (i.e.\ the stellar population
parameters) found both in the cluster sample and the cluster outskirts
sample. Finally, the intrinsic scatter of the parameter distributions
in the cluster sample will be estimated numerically and, combined with
the results from parameter--$\sigma$ relations, the degree to which
mass and SF time-scale influence the stellar populations of cluster
early-type galaxies will be determined. These investigations will
provide strong constraints for models of galaxy formation and
evolution.\\

The authors thank the referee, Reynier Peletier, for a careful reading
of the paper and for the many helpful suggestions that improved it
considerably.


\appendix

\section{Galaxy Catalogues}\label{gal cats}

In this appendix we give the details, line-strengths and stellar
population parameters of the galaxies observed as part of this
project. Table \ref{table_a1} gives the details of the galaxies such
as redshift and velocity dispersion. Column 5 denotes whether the
galaxy is considered an early-type galaxy or not, based on H$\alpha$
and/or [OIII]$\lambda$5007 \AA\ emission. Table \ref{table_a2}
presents their fully-corrected line-strengths while Table \ref{table_a3}
presents the stellar population parameters estimated from them.

Complete versions of these tables can be found on-line.

\begin{table*}
\caption{Details of the galaxies. Coordinates are in J2000.0;
  c$z_\odot$ and $\sigma$ units are $\kmpers$. The full version of
  this table is available electronically.\label{table_a1}}
\begin{tabular}{llcccccccrcrr}
\hline
Cluster  &  Name  &  R.A.  &  Dec.  &  ET  &  $\mathrm{m_{b_J}}$  &
$\mathrm{M_{b_J}}$  &  $\mathrm{m_{r_F}}$  &  $\mathrm{M_{r_F}}$  &
c$z_\odot$  &  $\epsilon_{\mathrm{c}z_\odot}$  &  $\sigma$  &
$\epsilon_\sigma$\\ 
\hline
A930  &  TGN215Z176  &  09 57 05.57  &  $-02$ 33 00.9  &  y  & 16.44  &  -20.88  &  15.29  &  -21.86  &  18071  &  80  &  242  & 14\\ 
&  TGN153Z276  &  09 58 05.30  &  $-04$ 13 09.6  &  y  & 16.10  &  -21.24  &  14.73  &  -22.43  &  19540  &  79  &  237  & 14\\ 
&  TGN153Z259  &  09 58 41.64  &  $-04$ 19 51.8  &  y  & 15.92  &  -21.35  &  14.68  &  -22.46  &  14955  &  79  &  201  & 15\\ 
&  TGN093Z199  &  09 59 44.18  &  $-05$ 22 01.2  &  y  & 15.42  &  -21.87  &  14.07  &  -23.07  &  15854  &  80  &  324  & 10\\ 
&  TGN093Z079  &  10 01 25.68  &  $-04$ 52 12.0  &  n  & 16.92  &  -20.38  &  15.63  &  -21.51  &  16873  &  78  &  104  & 18\\ 
&  TGN094Z319  &  10 01 40.38  &  $-05$ 20 56.5  &  n  & 17.00  &  -20.29  &  15.76  &  -21.38  &  15854  &  78  &  138  & 16\\ 
&  TGN217Z284  &  10 01 44.05  &  $-02$ 13 21.8  &  y  & 17.23  &  -20.09  &  15.98  &  -21.17  &  17982  &  78  &  119  & 17\\ 
&  TGN154Z324  &  10 02 15.49  &  $-04$ 44 41.7  &  n  & 16.70  &  -20.63  &  15.54  &  -21.61  &  19122  &  79  &  201  & 15\\ 
&  TGN094Z273  &  10 03 20.66  &  $-05$ 14 10.0  &  n  & 18.44  &  -18.89  &  17.16  &  -19.99  &  19194  &  80  &  122  & 19\\ 
&  TGN094Z252  &  10 03 37.28  &  $-05$ 06 01.2  &  n  & 15.90  &  -21.40  &  14.82  &  -22.32  &  16604  &  79  &  189  & 15\\ 
&  TGN094Z244  &  10 03 43.33  &  $-04$ 39 44.9  &  y  & 17.46  &  -19.87  &  16.07  &  -21.08  &  18972  &  79  &  202  & 15\\ 
&  TGN154Z183  &  10 03 52.99  &  $-04$ 08 43.9  &  n  & 17.09  &  -20.24  &  15.91  &  -21.24  &  18762  &  78  &  121  & 16\\ 
&  TGN154Z169  &  10 04 05.76  &  $-04$ 22 15.4  &  y  & 17.18  &  -20.15  &  16.24  &  -20.91  &  19062  &  79  &  174  & 17\\ 
&  TGN217Z071  &  10 04 11.62  &  $-02$ 51 12.8  &  y  & 16.65  &  -20.67  &  15.32  &  -21.83  &  18102  &  80  &  306  & 13\\ 
&  TGN095Z338  &  10 04 15.24  &  $-06$ 03 09.0  &  n  & 17.16  &  -20.16  &  15.93  &  -21.22  &  18474  &  79  &  137  & 13\\ 
&  TGN095Z321  &  10 04 49.94  &  $-05$ 11 22.6  &  n  & 17.43  &  -19.85  &  16.13  &  -21.01  &  15566  &  79  &  165  & 11\\ 
&  TGN094Z172  &  10 04 50.11  &  $-05$ 49 36.5  &  n  & 17.13  &  -20.19  &  15.88  &  -21.27  &  18624  &  78  &  135  & 14\\ 
&  TGN217Z056  &  10 04 53.59  &  $-03$ 25 50.1  &  n  & 17.31  &  -19.96  &  16.11  &  -21.02  &  14685  &  78  &  155  & 17\\ 
&  TGN094Z163  &  10 04 57.00  &  $-04$ 47 33.7  &  y  & 16.11  &  -21.22  &  14.79  &  -22.36  &  18822  &  80  &  289  & 11\\ 
&  TGN094Z153  &  10 04 58.31  &  $-05$ 03 09.9  &  y  & 16.81  &  -20.52  &  15.64  &  -21.51  &  18648  &  56  &  167  & 9\\ 
&  TGN095Z308  &  10 05 16.24  &  $-05$ 25 14.8  &  n  & 18.97  &  -18.31  &  17.67  &  -19.47  &  15536  &  78  &  112  & 14\\ 
&  TGN094Z130  &  10 05 22.24  &  $-06$ 06 35.1  &  n  & 16.78  &  -20.51  &  15.43  &  -21.71  &  16436  &  80  &  216  & 8\\ 
&  TGN094Z119  &  10 05 28.93  &  $-05$ 06 10.3  &  n  & 15.87  &  -21.42  &  14.53  &  -22.61  &  15771  &  56  &  231  & 8\\ 
&  TGN095Z275  &  10 06 03.80  &  $-05$ 27 34.8  &  n  & 15.68  &  -21.61  &  14.57  &  -22.57  &  15866  &  79  &  152  & 12\\ 
&  TGN037Z340  &  10 06 06.33  &  $-06$ 28 48.8  &  n  & 16.64  &  -20.65  &  15.27  &  -21.87  &  16436  &  80  &  220  & 8\\ 
&  TGN095Z262  &  10 06 28.17  &  $-05$ 30 08.1  &  n  & 17.35  &  -19.97  &  16.07  &  -21.08  &  18324  &  78  &   92  & 15\\ 
&  TGN154Z063  &  10 06 30.40  &  $-03$ 27 13.3  &  y  & 16.70  &  -20.62  &  15.51  &  -21.64  &  17953  &  79  &  206  & 15\\ 
&  TGN095Z259  &  10 06 31.11  &  $-06$ 16 45.9  &  y  & 18.52  &  -18.81  &  17.21  &  -19.94  &  18684  &  78  &   95  & 15\\ 
&  TGN217Z184  &  10 06 31.44  &  $-02$ 00 51.7  &  n  & 17.22  &  -20.12  &  15.94  &  -21.22  &  19721  &  79  &  250  & 13\\ 
&  TGN094Z077  &  10 06 35.21  &  $-05$ 52 01.0  &  y  & 17.35  &  -19.95  &  15.95  &  -21.20  &  17125  &  79  &  181  & 11\\ 
&  TGN095Z251  &  10 06 36.97  &  $-04$ 53 24.4  &  y  & 16.37  &  -20.91  &  15.03  &  -22.11  &  15506  &  79  &  223  & 10\\ 
&  TGN095Z242  &  10 06 39.54  &  $-05$ 38 36.3  &  n  & 16.41  &  -20.89  &  15.27  &  -21.87  &  16874  &  78  &  101  & 17\\ 
&  TGN095Z234  &  10 06 44.95  &  $-05$ 23 18.0  &  y  & 16.45  &  -20.87  &  15.08  &  -22.07  &  18523  &  80  &  314  & 13\\ 
&  TGN037Z036  &  10 06 45.99  &  $-06$ 18 25.0  &  n  & 17.58  &  -19.75  &  16.29  &  -20.86  &  19134  &  78  &  106  & 14\\ 
&  TGN217Z005  &  10 06 49.09  &  $-03$ 10 05.7  &  y  & 17.25  &  -20.09  &  15.94  &  -21.22  &  19422  &  80  &  271  & 14\\ 
&  TGN095Z220  &  10 06 52.30  &  $-05$ 33 13.0  &  n  & 17.29  &  -20.00  &  16.05  &  -21.09  &  16136  &  79  &  167  & 11\\ 
&  TGN037Z266  &  10 06 52.83  &  $-06$ 26 53.4  &  y  & 18.25  &  -19.07  &  16.90  &  -20.25  &  18084  &  78  &  104  & 14\\ 
&  TGN037Z263  &  10 06 55.53  &  $-06$ 30 41.6  &  y  & 19.37  &  -17.96  &  18.13  &  -19.02  &  19014  &  78  &  133  & 14\\ 
&  TGN094Z050  &  10 06 57.91  &  $-05$ 08 45.1  &  y  & 17.89  &  -19.44  &  16.59  &  -20.56  &  18714  &  79  &  191  & 12\\ 
&  TGN095Z217  &  10 06 57.93  &  $-05$ 47 55.2  &  n  & 16.77  &  -20.54  &  15.66  &  -21.49  &  17335  &  78  &  142  & 12\\ 
&  TGN095Z208  &  10 06 59.52  &  $-05$ 37 34.9  &  y  & 17.90  &  -19.41  &  16.57  &  -20.58  &  17485  &  79  &  160  & 11\\ 
&  TGN094Z043  &  10 07 00.48  &  $-05$ 20 21.5  &  y  & 18.47  &  -18.85  &  17.13  &  -20.02  &  18354  &  78  &  113  & 14\\ 
&  TGN095Z357  &  10 07 05.82  &  $-06$ 23 50.1  &  y  & 17.98  &  -19.34  &  16.63  &  -20.52  &  18534  &  79  &  181  & 11\\ 
&  TGN095Z200  &  10 07 06.60  &  $-05$ 36 11.0  &  y  & 19.26  &  -18.04  &  17.94  &  -19.20  &  16526  &  78  &   86  & 16\\ 
&  TGN094Z031  &  10 07 10.79  &  $-05$ 25 23.4  &  n  & 17.27  &  -20.05  &  15.95  &  -21.20  &  18114  &  79  &  156  & 11\\ 
&  TGN094Z023  &  10 07 20.36  &  $-05$ 17 24.3  &  n  & 17.80  &  -19.52  &  16.53  &  -20.62  &  18054  &  78  &   96  & 15\\ 
&  TGN095Z191  &  10 07 20.83  &  $-05$ 32 59.7  &  n  & 16.03  &  -21.27  &  14.73  &  -22.41  &  16705  &  80  &  251  & 12\\ 
&  TGN037Z015  &  10 07 27.54  &  $-06$ 08 48.5  &  n  & 17.35  &  -19.98  &  16.03  &  -21.12  &  19194  &  80  &  247  & 11\\ 
&  TGN095Z186  &  10 07 27.72  &  $-05$ 27 58.7  &  y  & 17.64  &  -19.65  &  16.40  &  -20.74  &  16466  &  79  &  186  & 12\\ 
&  TGN095Z177  &  10 07 30.20  &  $-04$ 45 01.9  &  y  & 16.85  &  -20.48  &  15.86  &  -21.29  &  18798  &  56  &  174  & 9\\ 
&  TGN095Z183  &  10 07 31.41  &  $-06$ 02 28.4  &  n  & 16.66  &  -20.64  &  15.39  &  -21.75  &  16556  &  79  &  171  & 11\\ 
&  TGN096Z255  &  10 07 31.56  &  $-05$ 58 42.1  &  n  & 17.91  &  -19.41  &  16.61  &  -20.54  &  18624  &  78  &  116  & 14\\ 
&  TGN155Z244  &  10 07 33.82  &  $-04$ 28 07.6  &  n  & 16.35  &  -20.97  &  15.11  &  -22.04  &  18523  &  79  &  192  & 15\\ 
&  TGN094Z002  &  10 07 34.31  &  $-05$ 20 05.4  &  y  & 16.70  &  -20.60  &  15.63  &  -21.51  &  16945  &  78  &  106  & 14\\ 
&  TGN096Z245  &  10 07 42.55  &  $-05$ 34 34.9  &  n  & 16.97  &  -20.33  &  15.61  &  -21.53  &  16714  &  56  &  205  & 9\\ 
&  TGN095Z154  &  10 07 53.06  &  $-04$ 39 28.0  &  y  & 16.93  &  -20.38  &  15.58  &  -21.57  &  17584  &  56  &  182  & 9\\ 
&  TGN095Z149  &  10 07 55.70  &  $-04$ 57 31.9  &  y  & 17.86  &  -19.46  &  16.49  &  -20.66  &  18204  &  79  &  169  & 12\\ 
&  TGN095Z147  &  10 08 02.07  &  $-06$ 01 51.3  &  y  & 17.23  &  -20.09  &  15.90  &  -21.25  &  18624  &  79  &  197  & 10\\ 
&  TGN095Z148  &  10 08 02.69  &  $-06$ 03 12.1  &  n  & 17.52  &  -19.80  &  16.27  &  -20.88  &  18444  &  78  &  103  & 15\\ 
&  TGN096Z221  &  10 08 05.17  &  $-05$ 02 16.4  &  n  & 16.97  &  -20.35  &  15.67  &  -21.48  &  18504  &  82  &  183  & 22\\ 
&  TGN096Z218  &  10 08 10.96  &  $-05$ 01 34.4  &  y  & 16.53  &  -20.79  &  15.12  &  -22.03  &  18468  &  57  &  284  & 8\\  
&  TGN095Z135  &  10 08 12.81  &  $-06$ 01 40.5  &  n  & 17.63  &  -19.70  &  16.46  &  -20.69  &  18894  &  78  &  118  & 14\\ 
\hline
\end{tabular}
\end{table*}

\begin{table*}
\vbox to220mm{\vfil Landscape table to go here.
  \caption{Line-strengths of the galaxies. Mg$_1$ and
    Mg$_2$ are in magnitudes all other indices are in \AA. The full
    version of this table is available
    electronically.\label{table_a2}} \vfil}
\end{table*}

\begin{table*}
  \caption{Stellar population parameters of the galaxies. Age
    units are Gyr. $N$ is the number of indices used to estimate the
    parameters. The full version of this table is available
    electronically.\label{table_a3}}
\begin{tabular}{lr@{.}lccr@{.}lcr@{.}lcc}
\hline
  Name & \multicolumn{2}{c}{Age}
  & $\epsilon^+_\mathrm{age}$ & $\epsilon^-_\mathrm{age}$ &
  \multicolumn{2}{c}{[Z/H]} & $\epsilon_\mathrm{[Z/H]}$ &
  \multicolumn{2}{c}{[$\alpha$/Fe]} &
  $\epsilon_{\mathrm{[}\alpha\mathrm{/Fe]}}$ & $N$\\
\hline
TGN215Z176   &     5 & 0 &   6.9 &  2.9 &   0 & 07 &  0.20 &   0 & 50 &  0.36 &    3\\
TGN153Z276   &        5 & 5 &   4.9 &   2.6 &   0 & 43 &   0.14 &   0 & 38 &  0.14 &    4\\
TGN153Z259   &       15 & 0 &   0.0 &   3.1 &   0 & 19 &   0.12 &  -0 & 19 &  0.09 &    3\\
TGN093Z199   &        5 & 0 &   4.1 &   1.4 &   0 & 57 &   0.08 &   0 & 04 &  0.13 &    4\\
TGN093Z079 & \multicolumn{2}{c}{---} &  --- &  --- & \multicolumn{2}{c}{---} &  --- & \multicolumn{2}{c}{---} &   ---  & --- \\
TGN094Z319 & \multicolumn{2}{c}{---} &  --- &  --- & \multicolumn{2}{c}{---} &  --- & \multicolumn{2}{c}{---} &   ---  & ---\\
TGN217Z284   &       15 & 0 &   0.0 &   4.1 &  -0 & 13 &   0.14 &  -0 & 19 &  0.16 &    3\\
TGN154Z324 & \multicolumn{2}{c}{---} &  --- &  --- & \multicolumn{2}{c}{---} &  --- & \multicolumn{2}{c}{---} &   ---  & ---\\
TGN094Z273 & \multicolumn{2}{c}{---} &  --- &  --- & \multicolumn{2}{c}{---} &  --- & \multicolumn{2}{c}{---} &   ---  & ---\\
TGN094Z252 & \multicolumn{2}{c}{---} &  --- &  --- & \multicolumn{2}{c}{---} &  --- & \multicolumn{2}{c}{---} &   ---  & ---\\
TGN094Z244   &        7 & 5 &   7.5 &   4.7 &   0 & 33 &   0.20 &   0 & 04 &  0.31 &    3\\
TGN154Z183 & \multicolumn{2}{c}{---} &  --- &  --- & \multicolumn{2}{c}{---} &  --- & \multicolumn{2}{c}{---} &   ---  & ---\\
TGN154Z169 & \multicolumn{2}{c}{---} &  --- &  --- & \multicolumn{2}{c}{---} &  --- & \multicolumn{2}{c}{---} &   ---  & ---\\
TGN217Z071   &        4 & 8 &   7.2 &   2.6 &   0 & 05 &   0.22 &   0 & 16 &  0.34 &    3\\
TGN095Z338 & \multicolumn{2}{c}{---} &  --- &  --- & \multicolumn{2}{c}{---} &  --- & \multicolumn{2}{c}{---} &   ---  & ---\\
TGN095Z321 & \multicolumn{2}{c}{---} &  --- &  --- & \multicolumn{2}{c}{---} &  --- & \multicolumn{2}{c}{---} &   ---  & ---\\
TGN094Z172 & \multicolumn{2}{c}{---} &  --- &  --- & \multicolumn{2}{c}{---} &  --- & \multicolumn{2}{c}{---} &   ---  & ---\\
TGN217Z056 & \multicolumn{2}{c}{---} &  --- &  --- & \multicolumn{2}{c}{---} &  --- & \multicolumn{2}{c}{---} &   ---  & ---\\
TGN094Z163   &        3 & 3 &   1.9 &   0.8 &   0 & 13 &   0.08 &   0 & 40 &  0.15 &    4\\
TGN094Z153 &       14 & 3 &   0.7 &   5.3 &   0 & 19 &  0.12 &   0 & 20 &  0.11 &    5\\
TGN095Z308 & \multicolumn{2}{c}{---} &  --- &  --- & \multicolumn{2}{c}{---} &  --- & \multicolumn{2}{c}{---} &   ---  & ---\\
TGN094Z130 & \multicolumn{2}{c}{---} &  --- &  --- & \multicolumn{2}{c}{---} &  --- & \multicolumn{2}{c}{---} &   ---  & ---\\
TGN094Z119 & \multicolumn{2}{c}{---} &  --- &  --- & \multicolumn{2}{c}{---} &  --- & \multicolumn{2}{c}{---} &   ---  & ---\\
TGN095Z275 & \multicolumn{2}{c}{---} &  --- &  --- & \multicolumn{2}{c}{---} &  --- & \multicolumn{2}{c}{---} &   ---  & ---\\
TGN037Z340 & \multicolumn{2}{c}{---} &  --- &  --- & \multicolumn{2}{c}{---} &  --- & \multicolumn{2}{c}{---} &   ---  & ---\\
TGN095Z262 & \multicolumn{2}{c}{---} &  --- &  --- & \multicolumn{2}{c}{---} &  --- & \multicolumn{2}{c}{---} &   ---  & ---\\
TGN154Z063   &       15 & 0 &   0.0 &   7.1 &  -0 & 13 &   0.16 &   0 & 33 &  0.31 &    3\\
TGN095Z259 &        2 & 7 &   3.0 &   1.1 &   0 & 27 &  0.18 &   0 & 41 &  0.15 &    5\\
TGN217Z184 & \multicolumn{2}{c}{---} &  --- &  --- & \multicolumn{2}{c}{---} &  --- & \multicolumn{2}{c}{---} &   ---  & ---\\
TGN094Z077 &       11 & 9 &   3.1 &   3.3 &   0 & 25 &  0.08 &   0 & 24 &  0.08 &    6\\
TGN095Z251   &        6 & 0 &   3.1 &   2.2 &   0 & 31 &   0.08 &   0 & 30 &  0.09 &    5\\
TGN095Z242 & \multicolumn{2}{c}{---} &  --- &  --- & \multicolumn{2}{c}{---} &  --- & \multicolumn{2}{c}{---} &   ---  & ---\\
TGN095Z234 & \multicolumn{2}{c}{---} &  --- &  --- & \multicolumn{2}{c}{---} &  --- & \multicolumn{2}{c}{---} &   ---  & ---\\
TGN037Z036 & \multicolumn{2}{c}{---} &  --- &  --- & \multicolumn{2}{c}{---} &  --- & \multicolumn{2}{c}{---} &   ---  & ---\\
TGN217Z005   &        6 & 9 &   6.2 &   3.4 &   0 & 03 &   0.14 &   0 & 50 &  0.31 &    3\\
TGN095Z220 & \multicolumn{2}{c}{---} &  --- &  --- & \multicolumn{2}{c}{---} &  --- & \multicolumn{2}{c}{---} &   ---  & ---\\
TGN037Z266 &        1 & 1 &   0.3 &   0.1 &   0 & 51 &  0.20 &   0 & 22 &  0.13 &    5\\
TGN037Z263   &       10 & 4 &   4.6 &   6.4 &   0 & 13 &   0.22 &  -0 & 00 &  0.32 &    3\\
TGN094Z050 &        9 & 1 &   4.6 &   3.8 &   0 & 21 &  0.12 &   0 & 02 &  0.12 &    5\\
TGN095Z217 & \multicolumn{2}{c}{---} &  --- &  --- & \multicolumn{2}{c}{---} &  --- & \multicolumn{2}{c}{---} &   ---  & ---\\
GN095Z208 &        6 & 3 &   3.7 &   1.9 &   0 & 19 &  0.08 &   0 & 37 &  0.09 &    5\\
TGN094Z043 &        5 & 0 &   6.4 &   2.6 &   0 & 11 &  0.16 &   0 & 29 &  0.16 &    5\\
TGN095Z357 &        3 & 3 &   2.4 &   0.8 &   0 & 55 &  0.12 &   0 & 39 &  0.10 &    5\\
TGN095Z200 &        7 & 9 &   6.4 &   4.6 &  -0 & 05 &  0.16 &   0 & 39 &  0.16 &    5\\
TGN094Z031 & \multicolumn{2}{c}{---} &  --- &  --- & \multicolumn{2}{c}{---} &  --- & \multicolumn{2}{c}{---} &   ---  & ---\\
TGN094Z023 & \multicolumn{2}{c}{---} &  --- &  --- & \multicolumn{2}{c}{---} &  --- & \multicolumn{2}{c}{---} &   ---  & ---\\
TGN095Z191 & \multicolumn{2}{c}{---} &  --- &  --- & \multicolumn{2}{c}{---} &  --- & \multicolumn{2}{c}{---} &   ---  & ---\\
TGN037Z015 & \multicolumn{2}{c}{---} &  --- &  --- & \multicolumn{2}{c}{---} &  --- & \multicolumn{2}{c}{---} &   ---  & ---\\
TGN095Z186 &        6 & 6 &   4.8 &   2.0 &   0 & 19 &  0.10 &   0 & 31 &  0.09 &    6\\
TGN095Z177 &        4 & 5 &   2.0 &   1.4 &   0 & 31 &  0.06 &   0 & 25 &  0.08 &    6\\
TGN095Z183 & \multicolumn{2}{c}{---} &  --- &  --- & \multicolumn{2}{c}{---} &  --- & \multicolumn{2}{c}{---} &   ---  & ---\\
TGN096Z255 & \multicolumn{2}{c}{---} &  --- &  --- & \multicolumn{2}{c}{---} &  --- & \multicolumn{2}{c}{---} &   ---  & ---\\
TGN155Z244 & \multicolumn{2}{c}{---} &  --- &  --- & \multicolumn{2}{c}{---} &  --- & \multicolumn{2}{c}{---} &   ---  & ---\\
TGN094Z002 &        1 & 9 &   0.6 &   0.3 &   0 & 23 &  0.08 &   0 & 25 &  0.09 &    6\\
TGN096Z245 & \multicolumn{2}{c}{---} &  --- &  --- & \multicolumn{2}{c}{---} &  --- & \multicolumn{2}{c}{---} &   ---  & ---\\
TGN095Z154 &        6 & 9 &   3.0 &   2.5 &   0 & 29 &  0.08 &   0 & 10 &  0.09 &    5\\
TGN095Z149 &        5 & 0 &   3.7 &   2.1 &   0 & 27 &  0.12 &   0 & 27 &  0.10 &    5\\
TGN095Z147 &        2 & 0 &   0.6 &   0.2 &   0 & 29 &  0.06 &   0 & 20 &  0.06 &    6\\
TGN095Z148 & \multicolumn{2}{c}{---} &  --- &  --- & \multicolumn{2}{c}{---} &  --- & \multicolumn{2}{c}{---} &   ---  & ---\\
TGN096Z221 & \multicolumn{2}{c}{---} &  --- &  --- & \multicolumn{2}{c}{---} &  --- & \multicolumn{2}{c}{---} &   ---  & ---\\
TGN096Z218 &       14 & 3 &   0.7 &   2.9 &   0 & 21 &  0.06 &   0 & 50 &  0.03 &    6\\
TGN095Z135 & \multicolumn{2}{c}{---} &  --- &  --- & \multicolumn{2}{c}{---} &  --- & \multicolumn{2}{c}{---} &   ---  & ---\\
\hline
\end{tabular}
\end{table*}

\bsp

\label{lastpage}


\begin{thebibliography}{99}
\bibitem[\protect\citeauthoryear{Bardelli et al.}{1994}]{bardelli94}
  Bardelli S., Zucca E., Vettolani G., Zamorani G., Scaramella R.,
  Collins C.~A., MacGillivray H.~T., 1994, MNRAS, 267, 665
\bibitem[\protect\citeauthoryear{Bardelli et al.}{1998}]{bardelli98}
                  Bardelli S., Zucca E., Zamorani G., Vettolani G.,
                  Scaramella R., 1998, MNRAS, 296, 599
\bibitem[\protect\citeauthoryear{Blair \& Gilmore}{1982}]{blair82}
  Blair M., Gilmore G., 1982, PASP, 94, 742
\bibitem[\protect\citeauthoryear{Borges et al.}{1995}]{borges95}
  Borges A.~C., Idiart T.~P., de Freitas Pacheco J.~A., Thevenin F.,
  1995, AJ, 110, 2408 
\bibitem[\protect\citeauthoryear{Burstein et al.}{1984}]{burstein84}
                  Burstein D., Faber S.~M., Gaskell C.~M., Krumm N.,
                  1984, ApJ, 287, 586
\bibitem[\protect\citeauthoryear{Burstein, Faber \& Gonzalez}{Burstein
  et al.}{1986}]{burstein86} Burstein D., Faber S.~M., Gonzalez J.~J.,
  1986, AJ, 91, 1130
\bibitem[\protect\citeauthoryear{Cardiel et al.}{1998}]{cardiel98}
  Cardiel N., Gorgas J., Cenarro J., Gonzalez J.~J., 1998, A\&AS, 127,
  597
\bibitem[\protect\citeauthoryear{Colless et al.}{2001}]{colless01}
  Colless M., Dalton G., Maddox S. et al., 2001, MNRAS, 328, 1039 
\bibitem[\protect\citeauthoryear{Colless et al.}{2003}]{colless03}
  Colless, M., Peterson B.~A., Jackson C. et al., 2003, preprint
  (astro-ph/0306581)
\bibitem[\protect\citeauthoryear{Dale et al.}{1997}]{dale97} Dale
  D.~A., Giovanelli R., Haynes M.~P., Scodeggio M., Hardy E.,
  Campusano L.~E., 1997, AJ, 114, 455
\bibitem[\protect\citeauthoryear{Davies, Sadler \& Peletier}{Davies et
  al.}{1993}]{davies93} Davies R.~L., Sadler E.~M., Peletier R.~F.,
  1993, MNRAS, 262, 650
\bibitem[\protect\citeauthoryear{Faber et al.}{1985}]{faber85} Faber
  S.~M., Friel E.~D., Burstein D., Gaskell C.~M., 1985, ApJS, 57, 711
\bibitem[\protect\citeauthoryear{Fisher, Franx \& Illingworth}{Fisher
    et al.}{1995}]{fisher95} Fisher D., Franx M., Illingworth G.,
    1995, ApJ, 448, 119
\bibitem[\protect\citeauthoryear{Gonzalez}{1993}]{gonzalez93} Gonz{\'
  a}lez J.~J., 1993, PhD thesis, Univ.\ of California Santa Cruz 
\bibitem[\protect\citeauthoryear{Gorgas \&
    Efstathiou}{1987}]{gorgas87} Gorgas J., Efstathiou G.,
    1987, in de Zeeuw P.~T., ed, Proc.\ IAU Symp. 127, Structure and
    Dynamics of Elliptical Galaxies, p.\ 189
\bibitem[\protect\citeauthoryear{Gorgas et al.}{1993}]{gorgas93}
  Gorgas J., Faber S.~M., Burstein D., Gonzalez J.~J., Courteau S.,
  Prosser C., 1993, ApJS, 86, 153
\bibitem[\protect\citeauthoryear{Greggio}{1997}]{greggio97} Greggio
  L., 1997, MNRAS, 285, 151
\bibitem[\protect\citeauthoryear{Huchra, Vogeley \& Geller}{Huchra et
    al.}{1999}]{huchra99} Huchra J.~P., Vogeley M.~S., Geller M.~J.,
    1999, ApJS, 121, 287 
\bibitem[\protect\citeauthoryear{Jones et al.}{2004}]{jones04} Jones
  D.~H., Saunders W., Colless M. et al., 2004, MNRAS, 355, 747
\bibitem[\protect\citeauthoryear{J{\o}rgensen}{1997}]{jorgensen97}
  J{\o}rgensen I., 1997, MNRAS, 288, 161
\bibitem[\protect\citeauthoryear{J{\o}rgensen}{1999}]{jorgensen99a}
  J{\o}rgensen I., 1999, MNRAS, 306, 607
\bibitem[\protect\citeauthoryear{J{\o}rgensen, Franx \&
    Kjaergaard}{J{\o}rgensen et al.}{1995}]{jorgensen95} J{\o}rgensen
    I., Franx M., Kjaergaard P., 1995, MNRAS, 276, 1341
\bibitem[\protect\citeauthoryear{Kaldare et al.}{2003}]{kaldare03}
  Kaldare R., Colless M., Raychaudhury S., Peterson B.~A., 2003,
  MNRAS, 339, 652
\bibitem[\protect\citeauthoryear{Katgert et al.}{1998}]{katgert98}
                  Katgert P., Mazure A., den Hartog R., Adami C.,
                  Biviano A., Perea J., 1998, A\&AS, 129, 399
\bibitem[\protect\citeauthoryear{Kuntschner}{2000}]{kuntschner00}
  Kuntschner H., 2000, MNRAS, 315, 184
\bibitem[\protect\citeauthoryear{Kuntschner et
    al.}{2002}]{kuntschner02} Kuntschner H., Smith R.~J., Colless M.,
    Davies R.~L., Kaldare R., Vazdekis A., 2002, MNRAS, 337, 172
\bibitem[\protect\citeauthoryear{Maraston}{1998}]{maraston98} Maraston
  C., 1998, MNRAS, 300, 872
\bibitem[\protect\citeauthoryear{Maraston et al.}{2002}]{maraston02}
                  Maraston C., Kissler-Patig M., Brodie J., Barmby P.,
                  Huchra J., 2002, Ap\&SS, 281, 137
\bibitem[\protect\citeauthoryear{Mehlert et al.}{1998}]{mehlert98}
  Mehlert D., Saglia R.~P., Bender R., Wegner G., 1998, A\&A, 332, 33
\bibitem[\protect\citeauthoryear{Moore et al.}{2002}]{moore02} Moore
  S.~A.~W., Lucey J.~R., Kuntschner H., Colless M., 2002, MNRAS, 336, 382
\bibitem[\protect\citeauthoryear{Nelan et al.}{2005}]{nelan05} Nelan
  J.~E., Smith R.~J., Hudson M.~J., Wegner G.~A., Lucey J.~R., Moore
  S.~A.~W., Quinney S.~J., Suntzeff N.~B., 2005, ApJ, 632, 137
\bibitem[\protect\citeauthoryear{Peletier}{1989}]{peletier89} Peletier
  R.~F., 1989, PhD thesis, Univ.\ of Groningen
\bibitem[\protect\citeauthoryear{Press et al.}{1992}]{press92} Press
  W.~H., Teukolsky S.~A., Vetterling W.~T., Flannery B.~P., 1992,
  Numerical recipes in FORTRAN. The art of scientific
  computing. Cambridge Univ. Press.
\bibitem[\protect\citeauthoryear{Proctor, Forbes \& Beasley}{Proctor
    et al.}{2004}]{proctor04} Proctor R.~N., Forbes D.~A., Beasley
    M.~A., 2004, MNRAS, 355, 1327
\bibitem[\protect\citeauthoryear{Puzia et al.}{2002}]{puzia02} Puzia
  T.~H., Saglia R.~P., Kissler-Patig M., Maraston C., Greggio L.,
  Renzini A., Ortolani S., 2002, A\&A, 395, 45
\bibitem[\protect\citeauthoryear{Rabin}{1982}]{rabin82} Rabin D.,
  1982, ApJ, 261, 85
\bibitem[\protect\citeauthoryear{S\'{a}nchez-Bl\'{a}zquez et
  al.}{2006}]{sanchez06} S\'{a}nchez-Bl\'{a}zquez P., Gorgas J.,
  Cardiel N., Gonz\'{a}lez J.~J., 2006, A\&A, 457, 787
\bibitem[\protect\citeauthoryear{Sarzi et al.}{2006}]{sarzi06} Sarzi
  M., Falc\'{o}n-Barroso J., Davies R. et al., 2006, MNRAS, 366, 1151
\bibitem[\protect\citeauthoryear{Shectman et al.}{1996}]{shectman96}
  Shectman S.~A., Landy S.~D., Oemler A., Tucker D.~L., Lin H.,
  Kirshner R.~P., Schechter P.~L., 1996, ApJ, 470, 172
\bibitem[\protect\citeauthoryear{Smith et al.}{2000}]{smith00} Smith
                  R.~J., Lucey J.~R., Hudson M.~J., Schlegel D.~J.,
                  Davies R.~L., 2000, MNRAS, 313, 469
\bibitem[\protect\citeauthoryear{Smith et al.}{2004}]{smith04} Smith
  R.~J., Hudson M.~J., Nelan J.~E. et al, 2004, AJ, 128, 1558
\bibitem[\protect\citeauthoryear{Teague, Carter \& Grey}{Teague et
    al.}{1990}]{teague90} Teague P.~F., Carter D., Gray P.~M., 1990,
    ApJS, 72, 715
\bibitem[\protect\citeauthoryear{Thomas, Maraston \& Bender}{Thomas et
    al.}{2003}]{thomas03a} Thomas D., Maraston C., Bender R., 2003,
    MNRAS, 339, 897
\bibitem[\protect\citeauthoryear{Trager et al.}{1998}]{trager98}
                  Trager S.~C., Worthey G., Faber S.~M., Burstein D.,
                  Gonzalez J.~J., 1998, ApJS, 116, 1
\bibitem[\protect\citeauthoryear{Trager et al.}{2000b}]{trager00b}
  Trager S.~C., Faber S.~M., Worthey G., Gonz{\' a}lez J.~J., 2000,
  AJ, 119, 1645
\bibitem[\protect\citeauthoryear{Trager et al.}{2000a}]{trager00a}
  Trager S.~C., Faber S.~M., Worthey G., Gonz{\' a}lez J.~J., 2000,
  AJ, 120, 165
\bibitem[\protect\citeauthoryear{Vazdekis et al.}{1996}]{vazdekis96}
  Vazdekis A., Casuso E., Peletier R.~F., Beckman J.~E., 1996, ApJS,
  106, 307
\bibitem[\protect\citeauthoryear{Vazdekis et al.}{2010}]{vazdekis10}
  Vazdekis A., S\'{a}nchez-Bl\'{a}zquez P., Falc\'{o}n-Barroso J.,
  Cenarro A.~J., Beasley M.~A., Cardiel N., Gorgas J., Peletier
  R. ~F., 2010, preprint (arXiv:1004.4439v1)
\bibitem[\protect\citeauthoryear{Wegner et al.}{1999}]{wegner99}
  Wegner G., Colless M., Saglia R.~P., McMahan R.~K., Davies R.~L.,
  Burstein D., Baggley G., 1999, MNRAS, 305, 259
\bibitem[\protect\citeauthoryear{Worthey}{1994}]{worthey94b} Worthey
  G., 1994, ApJS, 95, 107
\bibitem[\protect\citeauthoryear{Worthey}{1996}]{worthey96} Worthey
  G., 1996, in Leitherer C., Fritze-von-Alvensleben U., Huchra J.,
  eds, ASP Conf. Ser. Vol. 98, From Stars to Galaxies: the Impact of
  Stellar Physics on Galaxy Evolution, Astron. Soc. Pac., San
  Francisco, p. 467
\bibitem[\protect\citeauthoryear{Worthey \&
    Ottaviani}{1997}]{worthey97} Worthey G., Ottaviani D.~L., 1997,
    ApJS, 111, 377
\bibitem[\protect\citeauthoryear{Worthey, Faber \& Gonzalez}{Worthey
  et al.}{1992}]{worthey92a} Worthey G., Faber S.~M., Gonzalez J.~J.,
  1992, ApJ, 398, 69
\bibitem[\protect\citeauthoryear{Worthey et al.}{1994}]{worthey94a}
  Worthey G., Faber S.~M., Gonzalez J.~J., Burstein, D., 1994, ApJS,
  94, 687
\end{thebibliography}
\end{document}